\documentclass[reprint,prl,longbibliography,superscriptaddress,nofootinbib]{revtex4-2}
\usepackage{lipsum}
\usepackage{fancyhdr}
\usepackage{graphicx}
\usepackage[caption=false]{subfig}
\usepackage{braket}
\usepackage[dvipsnames]{xcolor}
\usepackage{mathtools}
\usepackage{amsmath}
\usepackage{amssymb}
\usepackage{hyperref}
\usepackage{esvect}
\usepackage{float}
\usepackage[normalem]{ulem}
\usepackage{notes2bib}
\graphicspath{{./Figures/}}

\hypersetup{
    colorlinks=true,
    linkcolor=blue,
    filecolor=magenta,      
    urlcolor=magenta,
    citecolor={blue},
    }

\newcommand{\edit}[1]{{\color{black} #1}}

\newcommand{\rereedit}[1]{{\color{black} #1}}
\newcommand{\rerereedit}[1]{{\color{black} #1}}

\allowdisplaybreaks

\begin{document}

\title{Improving On-Demand Single Photon Source Coherence and Indistinguishability Through a Time-Delayed Coherent Feedback}

\author{Gavin Crowder}
\email{gcrow088@uottawa.ca}
\affiliation{Department of Physics, University of Ottawa, Ottawa, Ontario, Canada, K1N 6N5}
\affiliation{Nexus for Quantum Technologies Institute, University of Ottawa, Ottawa, Ontario, Canada, K1N 6N5}
\affiliation{Department of Physics, Queen's University, Kingston, Ontario, Canada, K7L 3N6}
\author{Lora Ramunno}
\affiliation{Department of Physics, University of Ottawa, Ottawa, Ontario, Canada, K1N 6N5}
\affiliation{Nexus for Quantum Technologies Institute, University of Ottawa, Ottawa, Ontario, Canada, K1N 6N5}
\author{Stephen Hughes}
\affiliation{Department of Physics, Queen's University, Kingston, Ontario, Canada, K7L 3N6}

\begin{abstract}
Single photon sources (SPSs) are an essential resource for many quantum information technologies. We demonstrate how the inclusion of time-delayed coherent feedback in a scalable waveguide system, can significantly improve the two key SPS figures of merit: coherence and indistinguishability. Our feedback protocol is simulated using a quantum trajectory discretized waveguide model which can be used to directly model Hanbury Brown and Twiss (HBT) and Hong-Ou-Mandel (HOM) interferometers. With the proper choice of the round trip phase, the non-Markovian dynamics from the time-delayed feedback improves the indistinguishability of the SPS by up to \edit{57$\%$}. We also show how this mechanism suppresses the detrimental effects of off-chip decay and pure dephasing.
\end{abstract}
\maketitle

The reliable generation of single photons is important for the implementation of quantum information systems such as quantum cryptography and quantum computing \cite{Kiraz2004}. Semiconductor quantum dots are an excellent source for generating quantum light in the form of single photons or photon pairs \cite{Santori2004,Takemoto2004,Cui2006,He2013,Kalliakos2014,Paul2015,Hughes2019,He2019,Laferriere2022,Appel2022,Gines2022,Da_Lio2022}. In particular, it has been shown that upon excitation by an appropriate laser pulse, preparing the quantum dot in the excited state, the dot can emit a single photon into a desired output mode, e.g., a cavity or waveguide. This generation process is termed a ``on-demand'' due to the production of a photon number state after each single laser pulse \cite{Bracht2021}. 

The key figures of merit for on-demand single photon sources (SPSs) are the \edit{brightness}, $\eta$ (\edit{the average number of photons per laser pulse}), the second order coherence, $g^{(2)}(0)$ (related to the single photon purity), and the indistinguishability, $\mathcal{I}$ (a measure of the coherence of the emitted photon state) \cite{Fischer2016,Hughes2019}. Significant research efforts have been developed to improve the efficiency, purity, and coherence of SPS generation. State of the art systems have achieved $g^{(2)}(0) = 0.012$ and an indistinguishability of $\mathcal{I} = 0.962$ \cite{He2019}, though often these numbers are after filtering. Unwanted photon pair generation, out of plane emission, and dephasing events (such as through phonon absorption/emission \cite{Iles-Smith2017,Gustin2018b}) are all key challenges to overcome for continued improvement of these SPSs.

Feedback, where the output from a system is used as a stabilizing or control mechanism, has been well used across various platforms \cite{Dorner2002,Tufarelli2013,Carmele2013,Naumann2016,Lu2017,Pichler2017,Wiseman2006,Arias2020,Grigoletto2021,Hjelme1991,Franklin2014}. This is most often implemented through measurement-based feedback, where the output is measured to inform an external control which acts on the system \cite{Wiseman2006,Kubanek2009,Gillett2010,Arias2020,Rafiee2020,Grigoletto2021,Tan2020,Giovanni2021}. However, this approach is problematic for quantum information systems that rely on maintaining system coherence. Instead, feedback can be included at the system level and act back on the system itself to avoid measurement: {\it coherent feedback}. Recently, coherent feedback in waveguide quantum electrodynamic systems was shown to significantly alter the photon output statistics with continuous-wave pumping \cite{Dorner2002,Koshino2012,Tufarelli2013,Carmele2013,Hoi2015,Naumann2016,Lu2017,Pichler2017,Grimsmo2015,Kabuss2016,Pichler2016,Nemet2016,Hein2016,Guimond2016,Whalen2017,Naumann2017,Guimond2017,Forn2017,Whalen2019,Nemet2019,Calajo2019,Crowder2020,Harwood2021,Barkemeyer2021,Shi2021,Regidor2021a,Barkemeyer2022,Crowder2022}. \edit{Coherent delay loops have been used in previous experiments for generating quantum states of light such as cluster states \cite{Istrati2020,Steindl2021} or graph states \cite{Li2020}, by manipulating the output field of a SPS.} 

Here, we show how coherent feedback can be used in the pulsed laser regime to improve the figures of merit for a practical SPS. Accurate simulation of coherent feedback is a difficult challenge in quantum circuits, since it introduces a non-Markovian dynamic. To calculate $g^{(2)}(0)$ and $\mathcal{I}$ requires knowledge of the two-time correlation functions, which typically use the quantum regression theorem (QRT) (although other specialized approaches have been developed \edit{which avoid the QRT}, but they have their own practical limits \cite{Cosacchi2021,Richter2022}). The QRT relies on a Markov approximation which fails in the presence of coherent feedback and other specialized techniques must be used \cite{QuantumNoise,Stenius1996,Chruscinski2010}. \edit{To avoid the non-Markovian dynamic, collisional models can be used which include the waveguide at the system level, rather than treating it as a reservoir. We will use and extend a recently developed quantum trajectory discretized waveguide (QTDW) model \cite{Regidor2021a,Crowder2022} that combines a collisional model for the feedback section of the waveguide with the quantum trajectory formalism \cite{Tian1992,Dalibard1992}. Simulating individual realizations of the system gives insights into the underlying stochastic system dynamics and can be averaged to obtain the ensemble dynamics. Our model can also incorporate additional dephasing needed for simulating realistic quantum dots, a limitation of matrix product states, the other main technique for simulating feedback dynamics. Moreover, the QTDW model scales linearly with the number of photon states and is parallelizable.} 

In this Letter, we connect the QTDW model to photon correlation experiments, to compute $g^{(2)}(0)$ and $\mathcal{I}$ for waveguide quantum electrodynamic systems with coherent feedback, yielding direct simulations of Hanbury Brown and Twiss (HBT) or Hong-Ou-Mandel (HOM) interferometers. We summarize the key features of the QTDW model and outline its extension to simulate the HBT and HOM measurements. We then show how a coherent feedback can significantly improve the \edit{$\eta$}, $g^{(2)}(0)$, and $\mathcal{I}$ figures of merit for a quantum dot SPS by up to \edit{57$\%$}, and explain this improvement in the context of the feedback dynamics.

\begin{figure}
    \centering
\subfloat{%
  \includegraphics[width=0.94\columnwidth]{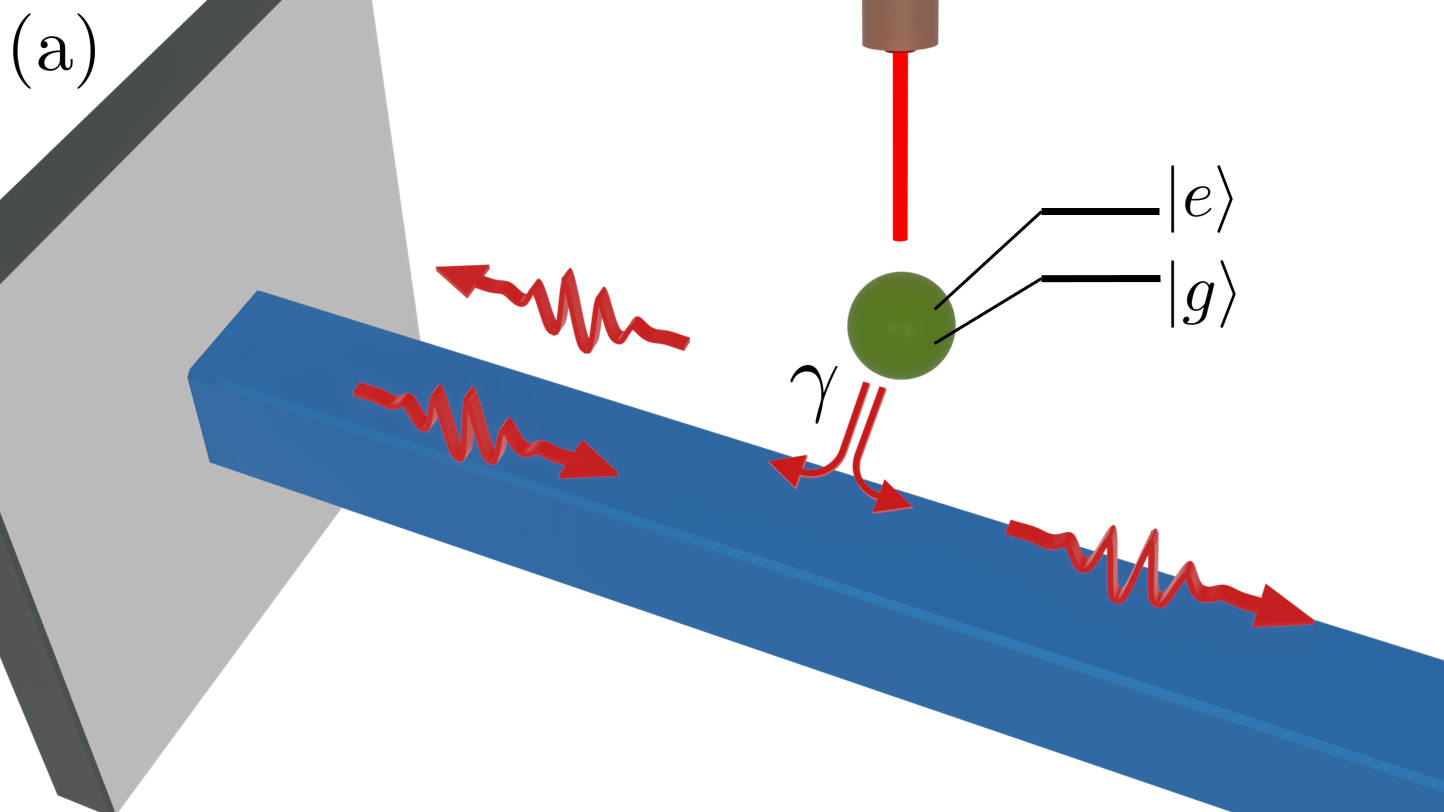}%
}
\vspace{-0.5cm}
\subfloat{%
  \includegraphics[width=0.4\columnwidth]{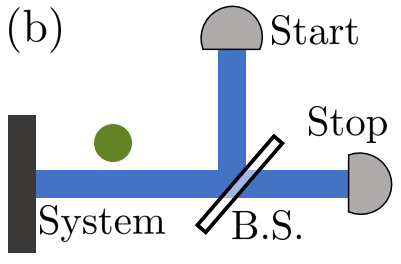}%
  \includegraphics[width=0.55\columnwidth]{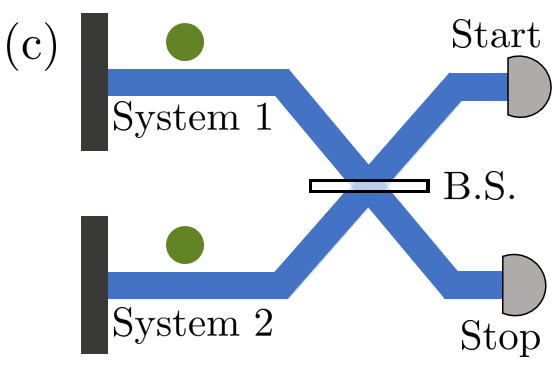}%
}
    \caption{Schematics of (a) the driven TLS setup with a coupled coherent feedback loop, (b) the HBT interferometer, and (c) the HOM interferometer; `B.S.' represents a beam splitter.}
    \label{Schematics}
\end{figure}

The system of interest, as shown in Fig.~\ref{Schematics}(a), is a single two-level system (TLS) coupled to a {\it truncated} waveguide which acts as a feedback loop, returning the output from the TLS to itself. An input laser pulse excites the TLS for SPS emission. The total Hamiltonian in natural units ($\hbar = 1$) for the system, waveguide, and pump (in the interaction picture \edit{of the laser} and with a rotating wave approximation) is $H = H_{\rm W} + H_{\rm TLS} + H_{\rm I} + H_{\rm pump}$. The TLS Hamiltonian is $H_{\rm TLS} = \delta \sigma^+ \sigma^-$ where $\sigma^+$ ($\sigma^-$) is the Pauli raising (lowering) operator and $\delta = \omega_0 - \omega_{\rm L}$ is the detuning between the TLS, with frequency $\omega_0$, and the laser, with frequency $\omega_L$. The pump Hamiltonian is $H_{\rm pump} = \Omega(t) /2 (\sigma^+ + \sigma^-)$ with the laser pulse shape given by $\Omega(t)$ which drives the quantum dot with a field orthogonal to the waveguide mode polarization \cite{Makhonin2014,Uppu2020,Dusanowski2022}.

In the continuous frequency domain, and in the \edit{same} interaction picture, the raising (lowering) operator for the photon modes in the waveguide are $b^{\dagger} (\omega)$ ($b(\omega)$) with Hamiltonian 
$    H_{\rm W} = \int_{-\infty}^{\infty} d\omega  \left(\omega {-} \omega_{\rm L} \right) b^\dagger (\omega) b (\omega)$, 
with 
$[ b (\omega), b^{\dagger} (\omega') ] = \delta(\omega{-}\omega')$.
The interaction between the waveguide and TLS, with total radiative rate $\gamma$ (and symmetric left and right coupling), is described by the Hamiltonian
$
    H_{\rm I} ={} 
    \int_{-\infty}^{\infty} d\omega \left[ \left( \sqrt{\frac{\gamma}{4\pi}} \sigma^+ b(\omega) \right. \right. 
    + \left. \left. \sqrt{\frac{\gamma}{4\pi}} e^{i(\phi_M + \omega \tau)} \sigma^+ b(\omega) \right) + \rm H.c. \right], \nonumber
$
which incorporates the influence of the feedback loop, including 
the round trip delay time, $\tau$, and a mirror phase change, $\phi_{\rm M}$.

To model the feedback loop, we use a QTDW technique \cite{Regidor2021a,Crowder2022}, where the field operators, \edit{$b(\omega)$}, for the waveguide section of interest ($-L < x < 0$) are transformed into the discrete time domain with operators $B_n$ \rerereedit{with commutation relation $[B_{n} , B^{\dagger}_{n'}] = \delta_{n,n'}$}. These operators describe a unidirectional slice of the waveguide field which travels around the feedback loop, interacting twice with the TLS, located at the beginning and end of the loop. \edit{Additional details on these time domain operators are given in the Supplementary Material~\cite{[{See Supplemental Material at }][{ for a discussion on the $B_n$ operators and a comparison of Markovian and non-Markovian feedback.}]supp}.} In this transformation, $H_{\rm TLS}$ and $H_{\rm pump}$ are unchanged and the interaction Hamiltonian becomes
\begin{equation}
    H_{\rm I} = \left ( \sqrt{\frac{\gamma} {2\Delta t}} \sigma^+ B_{N-1}  + e^{i \phi} \sqrt{\frac{\gamma} {2\Delta t}} \sigma^+ B_{0} \right) + \rm{H.c.},
\end{equation}
where $\phi = \phi_{\rm M} + \omega_0 \tau$ is the round trip phase change and $\Delta t = \tau / N$ is the time discretization of the feedback loop; now the TLS interacts with the incoming $(N-1)^{\rm th}$ bin ($B_{N-1}$) and the outgoing zeroth bin ($B_0$) which picks up the phase change $\phi$. The key advantage of this approach appears in the evolution of the $B_n$ operators under the waveguide Hamiltonian, which is
$U_{\rm W}^{\dagger}(\Delta t) B_{n} U_{\rm W} (\Delta t) = e^{-i \delta \Delta t} B_{n-1}$,
where $U_{\rm W} (\Delta t) = e^{-i H_{\rm W} \Delta t}$. This evolution acts to {\it pass} the waveguide field forward one bin each time step, where we assume linear dispersion. 

\edit{A limitation of this approach is that the total number of waveguide photons at any instant must be truncated. Typically, this is chosen to be two to capture the interference between multiple photons in the feedback loop and is a good approximation for most waveguide QED systems \cite{Regidor2021a}. For our system, where the ideal generation is a single photon per pulse, this truncation has no affect on the results.} The waveguide field outside of the loop, i.e., for $x > 0$, is modeled through the incoming and outgoing feedback loop bins, and the incoming bins are vacuum fields. During the quantum trajectory step, we take a simulated measurement of the outgoing bin ($B_0$) and project the full ket vector accordingly. After projection, there is no information in the outgoing bin and it can be dropped to retain a tractable Hilbert space (absorbing boundary condition). This is a stochastic process, but an average over many individual trajectories yields the ensemble average behavior of the system. 

Importantly (and a major advantage over \edit{matrix product states}) we also include two important dissipation channels in our model of the TLS: $(i)$ off chip decay from the TLS, $C_0 = \sqrt{\gamma_0} \sigma^-$, with rate $\gamma_0$, and (ii)  pure dephasing in the TLS, $C_1 = \sqrt{\gamma'} \sigma^+ \sigma^-$, with rate $\gamma'$. These are included as quantum jump operators following standard quantum trajectory theory \edit{\cite{Tian1992,Dalibard1992}}.

\edit{Conventionally, the brightness, $\eta$, is calculated as the total field emitted from the qubit into the waveguide from a single laser pulse. In the ensemble average picture this is calculated as $\eta = \int_0^{\infty} dt' \gamma \braket{\sigma^+ (t') \sigma^- (t')}$. In the quantum trajectory picture, the above expression can also be calculated using the average expectation of $\braket{\sigma^+ \sigma^-}$ across all trajectories. Equivalently 
since each trajectory contains a detection record of emitted photons from the waveguide, $\eta$ can be calculated by taking the average number of photons emitted in a trajectory.
}

Using the QTDW model, we take advantage of the natural link between the individual trajectories and experiments to calculate $g^{(2)}(0)$ and $\mathcal{I}$ through simulated HBT and HOM interferometers. While our technique can tractably calculate the two-time correlation functions, the HBT and HOM approaches are more intuitive and numerically much faster. In the HBT interferometer \cite{Hanbury1956}, see Fig.~\ref{Schematics}(b), the output field is sent to a 50:50 beam-splitter and entangled with the vacuum field. \rerereedit{The two new fields are then measured by two detectors, a {\it start} detector and a {\it stop} detector, with operators $B_{\rm{start}}^{\rm{HBT}} = B_0/\sqrt{2}$ and $B_{\rm{stop}}^{\rm{HBT}} = B_0/\sqrt{2}$ (with vacuum contributions not included as they do not play a role in the measurements)}. The detection times between the two detectors are correlated by taking the difference between the detection time on the {\it start} detector and all {\it stop} detection events, yielding a histogram of time-correlated detection records, $h_{\rm HBT}(t')$. The system is driven by consecutive pulses \edit{to build up a sufficient detection record, each} separated by a delay time, $T$, long enough for the system to completely relax into the ground state \edit{before sending in the next pulse}.

To simulate the HBT detection, we run each trajectory independently with a single pulse to drive the system and then allowing it to relax for a total run time of $T$. Next, we place these trajectories in sequential order so that each pulse is separated by $T$ and treat it as one long single quantum trajectory. Since the beam-splitter equally splits the output field, to get the time-correlated detection records we stochastically assign each output jump event in the trajectory to either the start or stop detector and then correlate the two detection records to build the histogram, $h_{\rm HBT} (t')$, where \rerereedit{$t'$} is the time difference between start and stop events. Assuming no correlations between pulses, when $T$ is chosen sufficiently large, then
\begin{equation}
    g^{(2)} (0) = \frac{A_{\rm HBT} (0)}{A_{\rm HBT} (T)} \left( 1 \pm \sqrt{\frac{1}{A_{\rm HBT} (0)} + \frac{1}{A_{\rm HBT} (T)}} \right),
\end{equation}
with $A_{\rm HBT} (0)$ the area of the peak centered at 0 and $A_{\rm HBT} (T)$ the area of the next peak, centered at $T$ \edit{\cite{Fischer2016}}. \edit{The central peak, $A_{\rm HBT} (0)$, represents the number of pairs of photons being emitted from the system and this is normalized by $A_{\rm HBT} (T)$, the photon count rate. Since each peak area is calculated as a count of events they carry an uncertainty of $\sqrt{N}$ where $N$ is the number of events. In these results a sufficiently large number of trajectories is carried out to approach an accurate average.}

In the HOM simulation \cite{Hong1987}, see Fig.~\ref{Schematics}(c), we take the interference between the output of two independent copies (`1' or `2') of our system incident on a 50:50 beam splitter (which can easily be extended to an unequal beam splitter). The output \rerereedit{bin operator} from system 1 (2) is \rerereedit{$B^{(1)}_0$ $(B^{(2)}_0)$. These are entangled on the beam splitter and sent on to the {\it start} and {\it stop} detectors, now with operators $B_{\rm{start}}^{\rm{HOM}} = {(B_0^{(1)} + B_0^{(2)})}/{\sqrt{2}}$ and $B_{\rm{stop}}^{\rm{HOM}} = {(B_0^{(1)} - B_0^{(2)})}/{\sqrt{2}}$}. We again drive both copies using consecutive pulses with delay time, $T$, and time-correlate the detection records of the two detectors. Similar to the HBT interferometer, the correlated detection record can be used to yield a histogram, $h_{\rm HOM} (t')$.

This HOM setup is more difficult to simulate with the quantum trajectory technique compared to the HBT interferometer, as the detection records are no longer accessible through our jump record of $B_0$. Instead we entangle two copies of our system using a tensor product, so that the total Hamiltonian is $H_{\rm HOM} = H^{(1)} \otimes H^{(2)}$, since the two systems are completely independent to begin with. Then, rather than taking our output operators to be $B^{(1)}_0$ and $B^{(2)}_0$, we use the entangled fields after the beam-splitter \rerereedit{$B_{\rm{start}}^{\rm{HOM}}$ and $B_{\rm{stop}}^{\rm{HOM}}$}. Next, when we detect a photon in the {\it start} ({\it stop}) detector, we apply the operator \rerereedit{$B_{\rm{start}}^{\rm{HOM}}$ ($B_{\rm{stop}}^{\rm{HOM}}$)} to the combined system. Detection times are recorded to build up detection records for the two detectors which we can time-correlate. \edit{Similarly to the HBT interferometer, f}rom $h_{\rm HOM} (t')$ one can calculate,
\begin{equation}
    g^{(2)}_{\rm HOM} (0) = \frac{A_{\rm HOM} (0)}{A_{\rm HOM} (T)} \left( 1 \pm \sqrt{\frac{1}{A_{\rm HOM} (0)} + \frac{1}{A_{\rm HOM} (T)}} \right),
\end{equation}
which is related to the indistinguishability through \cite{Fischer2016, Hughes2019}
\begin{equation}
    \mathcal{I} = 1 - g^{(2)}_{\rm HOM} (0).
\end{equation}

Without feedback, i.e., for a TLS coupled to an open waveguide \cite{Kalliakos2014,Hughes2019,Schnauber2019,Uppu2020}, to generate a SPS we pump the system with a $\pi$-pulse to excite the TLS and decay back to the ground state, emitting a single photon. \rereedit{Note that we correct for the bi-directional output in the results without feedback so both output fields are accounted for.} In the limit of an infinitely short pulse width, and neglecting other decoherence effects such as phonon interactions \cite{Hughes2019}, this TLS is a perfect SPS with $g^{(2)} (0) = 0$ and $\mathcal{I} = 1$. The perfection of these sources fail when a finite pulse width is taken to account (as this introduces a chance of two photon emission) and sources of detuning and off-chip decay are included (which spoils coherence) \cite{Gustin2018}. The TLS is driven by a Gaussian $\pi$-pulse, defined by 
$
    \Omega(t) = \left(\pi / \sigma \sqrt{2\pi}\right) {\rm exp} \left[ -t^2 / 2\sigma^2 \right],
$
where $\sigma = t_p/2\sqrt{2 \ln 2}$ and $t_p$ is the FWHM pulse width.

\begin{figure}
    \centering
    \includegraphics[width=0.99\columnwidth]{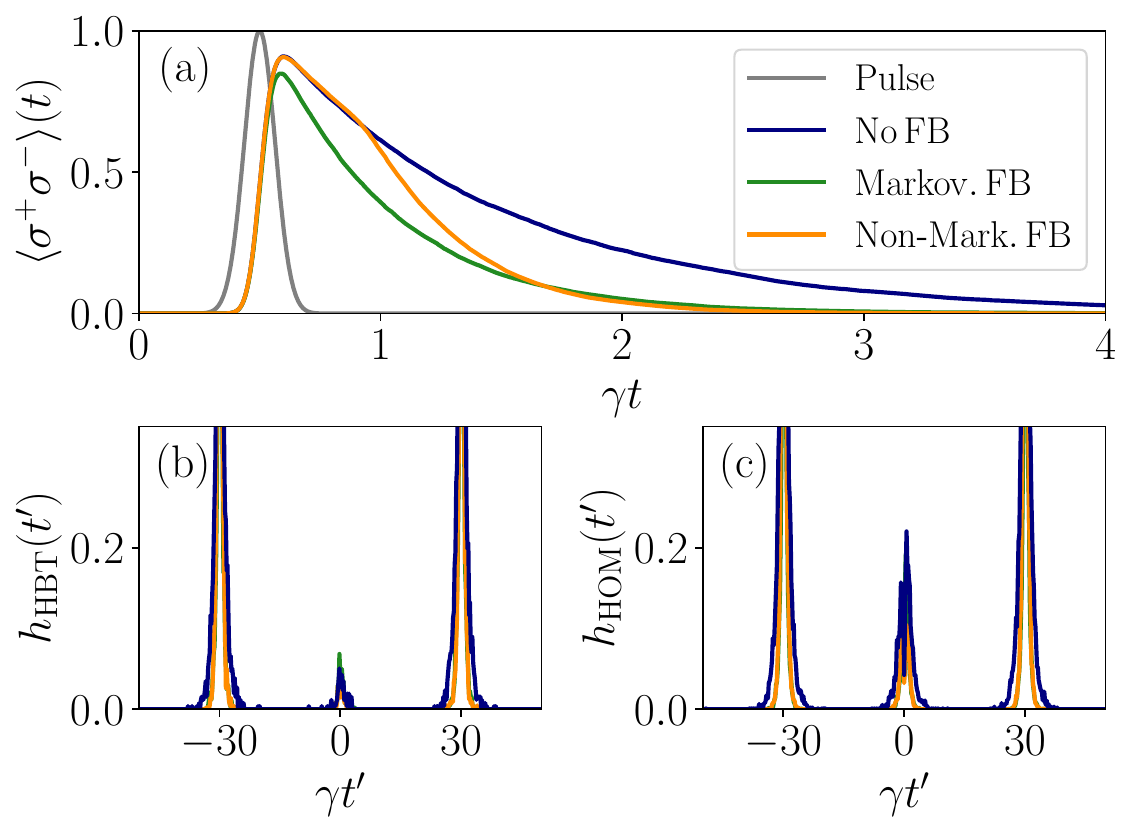}
    \vspace{-0.5cm}
    \caption{(a) The TLS population dynamics \rerereedit{without feedback, with Markovian feedback, and with non-Markovian feedback} for a pulse with width $\gamma t_p  = 0.15$. For this pulse width, the delay time used is $\gamma \tau = 3 \gamma t_p = 0.45$ so the \rerereedit{non-Markovian} feedback returns after the pulse. No additional loss channels are included. \rerereedit{In (b) and (c), the center peaks of the resulting HBT and HOM histograms for this feedback and pulse setup respectively}, with pure dephasing (using $\gamma' = \gamma$).
    }
    \label{Time_Dynamics}
\end{figure}

\edit{To include feedback, the delay time, $\tau$, and the phase change, $\phi$, must be strategically chosen to yield the best results. \rerereedit{We found this occurs when the non-Markovian feedback returns just after the pulse has finished driving the system, $\tau = 3 t_p$, and with a constructive interference, $\phi = 0$. The choice of delay time is discussed in the Supplementary Material~\cite{[{See Supplemental Material at }][{ for a discussion on the $B_n$ operators and a comparison of Markovian and non-Markovian feedback.}]supp}}. This works to limit both of the primary drawbacks to a perfect SPS: pair emission and additional loss channels. Pair emission occurs when the TLS emits a photon during the pulse and is re-excited, emitting a second photon. Thus, to minimize pair emission, a weak decay rate from the system during the pulse is best. However, in realistic systems, the TLS is limited by off-chip decay and pure dephasing. To minimize these, the TLS should decay as quickly as possible. Hence, a time-delayed coherent feedback works optimally to improve the SPS by not increasing the decay rate until after the pulse has finished interacting with the TLS. In Fig.~\ref{Time_Dynamics}(a), the TLS population dynamics ($\braket{\sigma^+ \sigma^-} (t)$) show how the delay in the \rerereedit{non-Markovian} feedback begins after the pulse has fully passed and the TLS decays much faster than the dynamics without feedback. This also yields the proper choice of $T$ to ensure the TLS has completely decayed into the ground state, which is taken to be $\gamma T = 20$ (well after the decay in Fig.~\ref{Time_Dynamics}(a)) in the HBT and HOM simulations. For the pulse widths in these results, $0.05 \leq \gamma t_p \leq 0.25$, this gives a range of delay times between $0.15 \leq \gamma \tau \leq 0.75$. This range is equivalent to a delay line with length 4.5 mm to 22.5 mm in a photonic crystal waveguide with $n_g = 10$ for a quantum dot system such as in \cite{2019Dan} with a decay rate of $\gamma \approx 1 {\rm{ns}}^{-1}$.}


\begin{figure*}
    \centering
    \includegraphics[width=1.9\columnwidth]{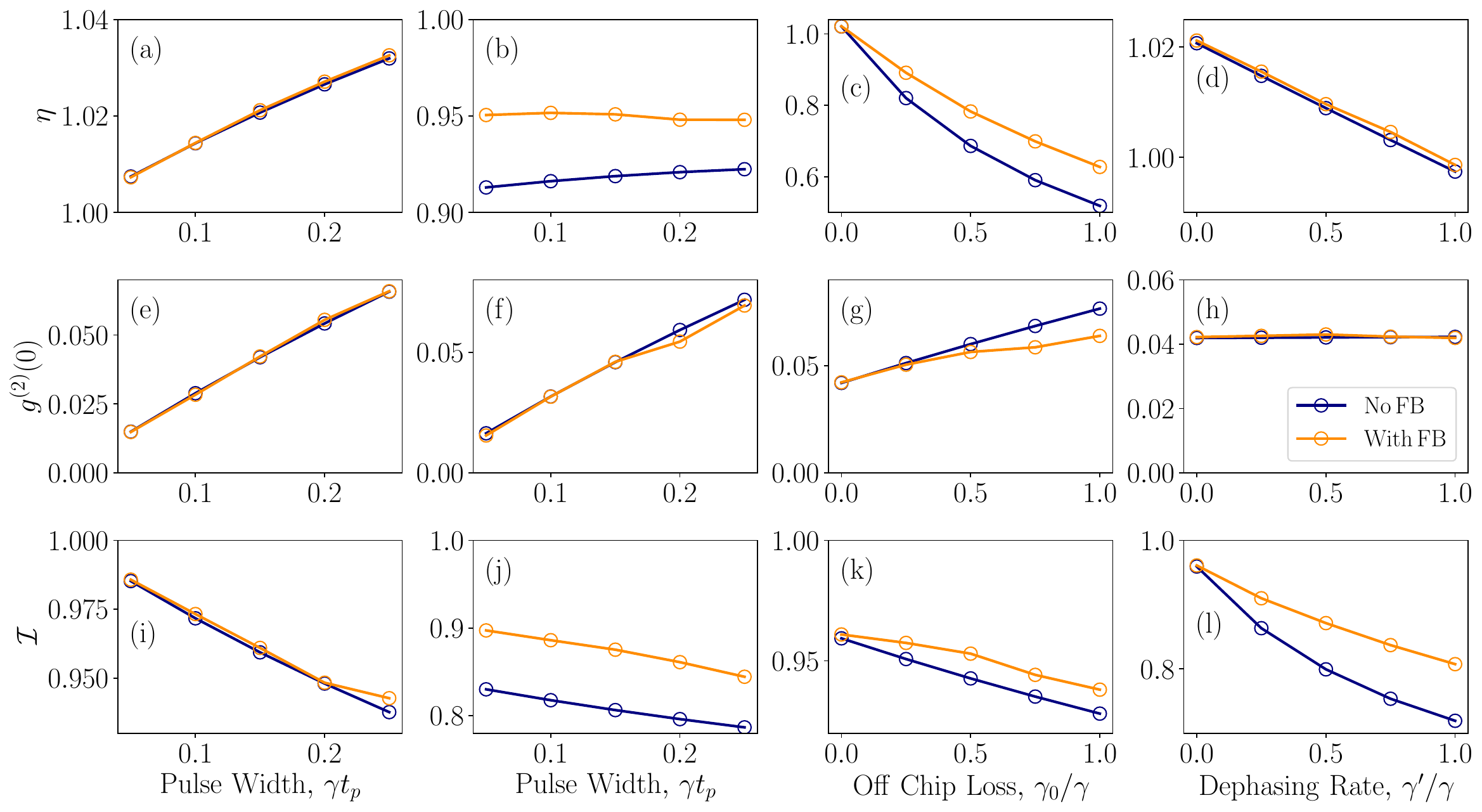}
    \vspace{-0.3cm}
    \caption{\edit{(a)-(d) The brightness, (e)-(h) the second order coherence, and (i)-(l) the indistinguishability of the output single photons. The width of the incoming pulse is varied without additional dissipation channels $C_0$ and $C_1$ in (a), (e), and (i), while these channels are included ($\gamma_0 = 0.1 \gamma$ and $\gamma' = 0.5 \gamma$) in (b), (f), and (j). The pulse width is then set at $\gamma t_p = 0.15$ and the off chip decay rate is varied in (c), (g), and (k) with $\gamma' = 0$, while the pure dephasing rate is varied in (d), (h), and (l) with $\gamma_0 = 0$. Each data point with feedback is an average of 100,000 trajectories for $\eta$ and $g^{(2)}(0)$, and 20,000 trajectories for $\mathcal{I}$.}
    }
    \label{CombinedScans}
\end{figure*}

Figure~\ref{CombinedScans} shows improvements in \edit{the brightness,} second order coherence, and indistinguishability when \rerereedit{non-Markovian} coherent feedback is included.  We display results with and without feedback as a function of changing pulse width, off chip decay rate, and pure dephasing rate.
\edit{In Figs.~\ref{CombinedScans}(a), (e), and (i) no additional loss channels are included so the only factor resulting in an imperfect SPS is pair emission. As shown in Fig.~\ref{Time_Dynamics}(a), the dynamics \rerereedit{without and with a non-Markovian feedback} have the same decay rate while the pulse drives the TLS so they undergo the same rate of pair emission. Thus, the SPS acts equally well with and without feedback.

When a realistic SPS is considered, i.e., with additional loss channels, the improvement with feedback is clearly seen. In Figs.~\ref{CombinedScans}(b), (f), and (j) we vary the pulse width but now have off-chip decay with rate $\gamma_0 = 0.1 \gamma$ and pure dephasing with rate $\gamma' = 0.5 \gamma$. The brightness is primarily affected by the off-chip loss as it introduces a new channel for the TLS to decay into, rather than the waveguide; thus, for some pulses, no corresponding photon detection is found. Across all pulse widths we find an average relative improvement of 39$\%$ in the brightness. The second order coherence is primarily affected by the pair emission from the system and so there is minimal relative improvement of 3.5$\%$ with feedback as the rates of pair emission are identical. Pure dephasing events spoil the coherence of the emitted photons and so have the largest affect on the indistinguishability. With feedback, the SPS has an average relative improvement of 35$\%$ in the indistinguishability.

A Markovian feedback, $\tau \rightarrow 0^+$, will also increase the spontaneous emission rate \rerereedit{at all times} when $\phi = 0$ \cite{Dorner2002} \rerereedit{as shown in Fig.~\ref{Time_Dynamics}(a)}. This will also improve the SPS performance in the presence of off-chip decay and pure dephasing, in an identical way as including a Purcell factor of 2. Crucially though, this also increases the pair emission rate which detrimentally affects the SPS. 
See \rerereedit{the Supplementary Material}~\cite{[{See Supplemental Material at }][{ for a discussion on the $B_n$ operators and a comparison of Markovian and non-Markovian feedback.}]supp}
for a comparison between the improvements with Markovian and non-Markovian feedback \rerereedit{where Markovian feedback produces a worse coherence and indistinguishability in all cases}. A similar effect in a coupled cavity-TLS setup has been shown in~\cite{Gustin2018b} where a non-Markovian Purcell factor can arise in the bad cavity regime with pulsed excitation.}

We also compare \edit{$\eta$}, $g^{(2)} (0)$ and $\mathcal{I}$ with and without feedback as a function of the off-chip decay rate in Figs.~\ref{CombinedScans}\edit{(c), (g), and (k),} and as a function of the pure dephasing rate in Figs.~\ref{CombinedScans}\edit{(d), (h), and (l),} with a fixed pulse width \edit{of $\gamma t_p = 0.15$}. We find that for off-chip decay (Figs.~\ref{CombinedScans}\edit{(c), (g), and (k)}), feedback does an excellent job of suppressing the effects of the additional output channel, decreasing the impact on \edit{brightness by 26$\%$,} coherence by \edit{39$\%$} and on indistinguishability by \edit{57$\%$}. For the pure dephasing rate  (Figs.~\ref{CombinedScans}\edit{(d), (h), and (l),}), \edit{$\eta$ and } $g^{(2)} (0)$ \edit{are} minimally affected by the increase in pure dephasing events so there is little difference between the results with and without feedback. Pure dephasing has its greatest affect on the indistinguishability, and feedback does a good job of reducing its impact, suppressing its effects by \edit{44$\%$}. \edit{Figure~\ref{Time_Dynamics}(c) shows an example HOM histogram when $\gamma' = \gamma$ highlighting the decreased area of the central peak in the case with \rerereedit{non-Markovian} feedback leading to the improvement in indistinguishability.}


In summary, we have exploited a QTDW model to calculate the $g^{(2)}(0)$ and $\mathcal{I}$ figures of merit for a SPS by directly simulating an HBT or HOM interferometer. This approach allows one to bypass the quantum regression theorem in order to simulate the non-Markovian dynamics that arise when coherent feedback is included in a waveguide quantum electrodynamic system. Using this technique, we have shown that direct inclusion of a time-delayed coherent feedback can improve the brightness, coherence, and indistinguishability of a SPS, even when pure dephasing is a significant source of decoherence in the system. This work harnesses a single characteristic of feedback, namely its ability to enhance spontaneous emission, in order to achieve these improvements.  In future work, other aspects of feedback such as excitation trapping and additional nonlinear behavior are also expected to be useful for photon pair emission and other quantum information system objectives.

\vspace{0.1cm}

\acknowledgements
This work was supported by the Natural Sciences and Engineering Research Council of Canada (NSERC), the National Research Council of Canada (NRC) through the Small Teams Ideation Program QPIC, Queen's University, and the University of Ottawa. We thank Dan Dalacu and Robin Williams for support and useful discussions.

\bibliography{PaperBib}

\begin{thebibliography}{74}%
\makeatletter
\providecommand \@ifxundefined [1]{%
 \@ifx{#1\undefined}
}%
\providecommand \@ifnum [1]{%
 \ifnum #1\expandafter \@firstoftwo
 \else \expandafter \@secondoftwo
 \fi
}%
\providecommand \@ifx [1]{%
 \ifx #1\expandafter \@firstoftwo
 \else \expandafter \@secondoftwo
 \fi
}%
\providecommand \natexlab [1]{#1}%
\providecommand \enquote  [1]{``#1''}%
\providecommand \bibnamefont  [1]{#1}%
\providecommand \bibfnamefont [1]{#1}%
\providecommand \citenamefont [1]{#1}%
\providecommand \href@noop [0]{\@secondoftwo}%
\providecommand \href [0]{\begingroup \@sanitize@url \@href}%
\providecommand \@href[1]{\@@startlink{#1}\@@href}%
\providecommand \@@href[1]{\endgroup#1\@@endlink}%
\providecommand \@sanitize@url [0]{\catcode `\\12\catcode `\$12\catcode
  `\&12\catcode `\#12\catcode `\^12\catcode `\_12\catcode `\%12\relax}%
\providecommand \@@startlink[1]{}%
\providecommand \@@endlink[0]{}%
\providecommand \url  [0]{\begingroup\@sanitize@url \@url }%
\providecommand \@url [1]{\endgroup\@href {#1}{\urlprefix }}%
\providecommand \urlprefix  [0]{URL }%
\providecommand \Eprint [0]{\href }%
\providecommand \doibase [0]{https://doi.org/}%
\providecommand \selectlanguage [0]{\@gobble}%
\providecommand \bibinfo  [0]{\@secondoftwo}%
\providecommand \bibfield  [0]{\@secondoftwo}%
\providecommand \translation [1]{[#1]}%
\providecommand \BibitemOpen [0]{}%
\providecommand \bibitemStop [0]{}%
\providecommand \bibitemNoStop [0]{.\EOS\space}%
\providecommand \EOS [0]{\spacefactor3000\relax}%
\providecommand \BibitemShut  [1]{\csname bibitem#1\endcsname}%
\let\auto@bib@innerbib\@empty
\bibitem [{\citenamefont {Kiraz}\ \emph {et~al.}(2004)\citenamefont {Kiraz},
  \citenamefont {Atatüre},\ and\ \citenamefont {Imamoğlu}}]{Kiraz2004}%
  \BibitemOpen
  \bibfield  {author} {\bibinfo {author} {\bibfnamefont {A.}~\bibnamefont
  {Kiraz}}, \bibinfo {author} {\bibfnamefont {M.}~\bibnamefont {Atatüre}},\
  and\ \bibinfo {author} {\bibfnamefont {A.}~\bibnamefont {Imamoğlu}},\
  }\bibfield  {title} {\bibinfo {title} {Quantum-dot single-photon sources:
  {Prospects} for applications in linear optics quantum-information
  processing},\ }\href {https://doi.org/10.1103/PhysRevA.69.032305} {\bibfield
  {journal} {\bibinfo  {journal} {Physical Review A}\ }\textbf {\bibinfo
  {volume} {69}},\ \bibinfo {pages} {032305} (\bibinfo {year}
  {2004})}\BibitemShut {NoStop}%
\bibitem [{\citenamefont {Santori}\ \emph {et~al.}(2004)\citenamefont
  {Santori}, \citenamefont {Fattal}, \citenamefont {Vučković}, \citenamefont
  {Solomon}, \citenamefont {Waks},\ and\ \citenamefont
  {Yamamoto}}]{Santori2004}%
  \BibitemOpen
  \bibfield  {author} {\bibinfo {author} {\bibfnamefont {C.}~\bibnamefont
  {Santori}}, \bibinfo {author} {\bibfnamefont {D.}~\bibnamefont {Fattal}},
  \bibinfo {author} {\bibfnamefont {J.}~\bibnamefont {Vučković}}, \bibinfo
  {author} {\bibfnamefont {G.~S.}\ \bibnamefont {Solomon}}, \bibinfo {author}
  {\bibfnamefont {E.}~\bibnamefont {Waks}},\ and\ \bibinfo {author}
  {\bibfnamefont {Y.}~\bibnamefont {Yamamoto}},\ }\bibfield  {title} {\bibinfo
  {title} {Submicrosecond correlations in photoluminescence from {InAs} quantum
  dots},\ }\href {https://doi.org/10.1103/PhysRevB.69.205324} {\bibfield
  {journal} {\bibinfo  {journal} {Physical Review B}\ }\textbf {\bibinfo
  {volume} {69}},\ \bibinfo {pages} {205324} (\bibinfo {year}
  {2004})}\BibitemShut {NoStop}%
\bibitem [{\citenamefont {Takemoto}\ \emph {et~al.}(2004)\citenamefont
  {Takemoto}, \citenamefont {Sakuma}, \citenamefont {Hirose}, \citenamefont
  {Usuki}, \citenamefont {Yokoyama}, \citenamefont {Miyazawa}, \citenamefont
  {Takatsu},\ and\ \citenamefont {Arakawa}}]{Takemoto2004}%
  \BibitemOpen
  \bibfield  {author} {\bibinfo {author} {\bibfnamefont {K.}~\bibnamefont
  {Takemoto}}, \bibinfo {author} {\bibfnamefont {Y.}~\bibnamefont {Sakuma}},
  \bibinfo {author} {\bibfnamefont {S.}~\bibnamefont {Hirose}}, \bibinfo
  {author} {\bibfnamefont {T.}~\bibnamefont {Usuki}}, \bibinfo {author}
  {\bibfnamefont {N.}~\bibnamefont {Yokoyama}}, \bibinfo {author}
  {\bibfnamefont {T.}~\bibnamefont {Miyazawa}}, \bibinfo {author}
  {\bibfnamefont {M.}~\bibnamefont {Takatsu}},\ and\ \bibinfo {author}
  {\bibfnamefont {Y.}~\bibnamefont {Arakawa}},\ }\bibfield  {title} {\bibinfo
  {title} {Non-classical {Photon} {Emission} from a {Single} {InAs}/{InP}
  {Quantum} {Dot} in the 1.3-µm {Optical}-{Fiber} {Band}},\ }\href
  {https://doi.org/10.1143/JJAP.43.L993} {\bibfield  {journal} {\bibinfo
  {journal} {Japanese Journal of Applied Physics}\ }\textbf {\bibinfo {volume}
  {43}},\ \bibinfo {pages} {L993} (\bibinfo {year} {2004})}\BibitemShut
  {NoStop}%
\bibitem [{\citenamefont {Cui}\ and\ \citenamefont {Raymer}(2006)}]{Cui2006}%
  \BibitemOpen
  \bibfield  {author} {\bibinfo {author} {\bibfnamefont {G.}~\bibnamefont
  {Cui}}\ and\ \bibinfo {author} {\bibfnamefont {M.~G.}\ \bibnamefont
  {Raymer}},\ }\bibfield  {title} {\bibinfo {title} {Emission spectra and
  quantum efficiency of single-photon sources in the cavity-{QED}
  strong-coupling regime},\ }\href {https://doi.org/10.1103/PhysRevA.73.053807}
  {\bibfield  {journal} {\bibinfo  {journal} {Physical Review A}\ }\textbf
  {\bibinfo {volume} {73}},\ \bibinfo {pages} {053807} (\bibinfo {year}
  {2006})}\BibitemShut {NoStop}%
\bibitem [{\citenamefont {He}\ \emph {et~al.}(2013)\citenamefont {He},
  \citenamefont {He}, \citenamefont {Wei}, \citenamefont {Wu}, \citenamefont
  {Atatüre}, \citenamefont {Schneider}, \citenamefont {H\"{o}fling},
  \citenamefont {Kamp}, \citenamefont {Lu},\ and\ \citenamefont
  {Pan}}]{He2013}%
  \BibitemOpen
  \bibfield  {author} {\bibinfo {author} {\bibfnamefont {Y.-M.}\ \bibnamefont
  {He}}, \bibinfo {author} {\bibfnamefont {Y.}~\bibnamefont {He}}, \bibinfo
  {author} {\bibfnamefont {Y.-J.}\ \bibnamefont {Wei}}, \bibinfo {author}
  {\bibfnamefont {D.}~\bibnamefont {Wu}}, \bibinfo {author} {\bibfnamefont
  {M.}~\bibnamefont {Atatüre}}, \bibinfo {author} {\bibfnamefont
  {C.}~\bibnamefont {Schneider}}, \bibinfo {author} {\bibfnamefont
  {S.}~\bibnamefont {H\"{o}fling}}, \bibinfo {author} {\bibfnamefont
  {M.}~\bibnamefont {Kamp}}, \bibinfo {author} {\bibfnamefont {C.-Y.}\
  \bibnamefont {Lu}},\ and\ \bibinfo {author} {\bibfnamefont {J.-W.}\
  \bibnamefont {Pan}},\ }\bibfield  {title} {\bibinfo {title} {On-demand
  semiconductor single-photon source with near-unity indistinguishability},\
  }\href {https://doi.org/10.1038/nnano.2012.262} {\bibfield  {journal}
  {\bibinfo  {journal} {Nature Nanotechnology}\ }\textbf {\bibinfo {volume}
  {8}},\ \bibinfo {pages} {213} (\bibinfo {year} {2013})}\BibitemShut {NoStop}%
\bibitem [{\citenamefont {Kalliakos}\ \emph {et~al.}(2014)\citenamefont
  {Kalliakos}, \citenamefont {Brody}, \citenamefont {Schwagmann}, \citenamefont
  {Bennett}, \citenamefont {Ward}, \citenamefont {Ellis}, \citenamefont
  {Skiba-Szymanska}, \citenamefont {Farrer}, \citenamefont {Griffiths},
  \citenamefont {Jones}, \citenamefont {Ritchie},\ and\ \citenamefont
  {Shields}}]{Kalliakos2014}%
  \BibitemOpen
  \bibfield  {author} {\bibinfo {author} {\bibfnamefont {S.}~\bibnamefont
  {Kalliakos}}, \bibinfo {author} {\bibfnamefont {Y.}~\bibnamefont {Brody}},
  \bibinfo {author} {\bibfnamefont {A.}~\bibnamefont {Schwagmann}}, \bibinfo
  {author} {\bibfnamefont {A.~J.}\ \bibnamefont {Bennett}}, \bibinfo {author}
  {\bibfnamefont {M.~B.}\ \bibnamefont {Ward}}, \bibinfo {author}
  {\bibfnamefont {D.~J.~P.}\ \bibnamefont {Ellis}}, \bibinfo {author}
  {\bibfnamefont {J.}~\bibnamefont {Skiba-Szymanska}}, \bibinfo {author}
  {\bibfnamefont {I.}~\bibnamefont {Farrer}}, \bibinfo {author} {\bibfnamefont
  {J.~P.}\ \bibnamefont {Griffiths}}, \bibinfo {author} {\bibfnamefont
  {G.~A.~C.}\ \bibnamefont {Jones}}, \bibinfo {author} {\bibfnamefont {D.~A.}\
  \bibnamefont {Ritchie}},\ and\ \bibinfo {author} {\bibfnamefont {A.~J.}\
  \bibnamefont {Shields}},\ }\bibfield  {title} {\bibinfo {title} {In-plane
  emission of indistinguishable photons generated by an integrated quantum
  emitter},\ }\href {https://doi.org/10.1063/1.4881887} {\bibfield  {journal}
  {\bibinfo  {journal} {Applied Physics Letters}\ }\textbf {\bibinfo {volume}
  {104}},\ \bibinfo {pages} {221109} (\bibinfo {year} {2014})},\ \bibinfo
  {note} {publisher: American Institute of Physics}\BibitemShut {NoStop}%
\bibitem [{\citenamefont {Paul}\ \emph {et~al.}(2015)\citenamefont {Paul},
  \citenamefont {Kettler}, \citenamefont {Zeuner}, \citenamefont {Clausen},
  \citenamefont {Jetter},\ and\ \citenamefont {Michler}}]{Paul2015}%
  \BibitemOpen
  \bibfield  {author} {\bibinfo {author} {\bibfnamefont {M.}~\bibnamefont
  {Paul}}, \bibinfo {author} {\bibfnamefont {J.}~\bibnamefont {Kettler}},
  \bibinfo {author} {\bibfnamefont {K.}~\bibnamefont {Zeuner}}, \bibinfo
  {author} {\bibfnamefont {C.}~\bibnamefont {Clausen}}, \bibinfo {author}
  {\bibfnamefont {M.}~\bibnamefont {Jetter}},\ and\ \bibinfo {author}
  {\bibfnamefont {P.}~\bibnamefont {Michler}},\ }\bibfield  {title} {\bibinfo
  {title} {Metal-organic vapor-phase epitaxy-grown ultra-low density
  {InGaAs}/{GaAs} quantum dots exhibiting cascaded single-photon emission at
  1.3 $\mu$m},\ }\href {https://doi.org/10.1063/1.4916349} {\bibfield
  {journal} {\bibinfo  {journal} {Applied Physics Letters}\ }\textbf {\bibinfo
  {volume} {106}},\ \bibinfo {pages} {122105} (\bibinfo {year}
  {2015})}\BibitemShut {NoStop}%
\bibitem [{\citenamefont {Hughes}\ \emph {et~al.}(2019)\citenamefont {Hughes},
  \citenamefont {Franke}, \citenamefont {Gustin}, \citenamefont
  {Kamandar~Dezfouli}, \citenamefont {Knorr},\ and\ \citenamefont
  {Richter}}]{Hughes2019}%
  \BibitemOpen
  \bibfield  {author} {\bibinfo {author} {\bibfnamefont {S.}~\bibnamefont
  {Hughes}}, \bibinfo {author} {\bibfnamefont {S.}~\bibnamefont {Franke}},
  \bibinfo {author} {\bibfnamefont {C.}~\bibnamefont {Gustin}}, \bibinfo
  {author} {\bibfnamefont {M.}~\bibnamefont {Kamandar~Dezfouli}}, \bibinfo
  {author} {\bibfnamefont {A.}~\bibnamefont {Knorr}},\ and\ \bibinfo {author}
  {\bibfnamefont {M.}~\bibnamefont {Richter}},\ }\bibfield  {title} {\bibinfo
  {title} {Theory and {Limits} of {On}-{Demand} {Single}-{Photon} {Sources}
  {Using} {Plasmonic} {Resonators}: {A} {Quantized} {Quasinormal} {Mode}
  {Approach}},\ }\href {https://doi.org/10.1021/acsphotonics.9b00849}
  {\bibfield  {journal} {\bibinfo  {journal} {ACS Photonics}\ }\textbf
  {\bibinfo {volume} {6}},\ \bibinfo {pages} {2168} (\bibinfo {year}
  {2019})}\BibitemShut {NoStop}%
\bibitem [{\citenamefont {He}\ \emph {et~al.}(2019)\citenamefont {He},
  \citenamefont {Wang}, \citenamefont {Wang}, \citenamefont {Chen},
  \citenamefont {Ding}, \citenamefont {Qin}, \citenamefont {Duan},
  \citenamefont {Chen}, \citenamefont {Li}, \citenamefont {Liu}, \citenamefont
  {Schneider}, \citenamefont {Atatüre}, \citenamefont {H\"{o}fling},
  \citenamefont {Lu},\ and\ \citenamefont {Pan}}]{He2019}%
  \BibitemOpen
  \bibfield  {author} {\bibinfo {author} {\bibfnamefont {Y.-M.}\ \bibnamefont
  {He}}, \bibinfo {author} {\bibfnamefont {H.}~\bibnamefont {Wang}}, \bibinfo
  {author} {\bibfnamefont {C.}~\bibnamefont {Wang}}, \bibinfo {author}
  {\bibfnamefont {M.-C.}\ \bibnamefont {Chen}}, \bibinfo {author}
  {\bibfnamefont {X.}~\bibnamefont {Ding}}, \bibinfo {author} {\bibfnamefont
  {J.}~\bibnamefont {Qin}}, \bibinfo {author} {\bibfnamefont {Z.-C.}\
  \bibnamefont {Duan}}, \bibinfo {author} {\bibfnamefont {S.}~\bibnamefont
  {Chen}}, \bibinfo {author} {\bibfnamefont {J.-P.}\ \bibnamefont {Li}},
  \bibinfo {author} {\bibfnamefont {R.-Z.}\ \bibnamefont {Liu}}, \bibinfo
  {author} {\bibfnamefont {C.}~\bibnamefont {Schneider}}, \bibinfo {author}
  {\bibfnamefont {M.}~\bibnamefont {Atatüre}}, \bibinfo {author}
  {\bibfnamefont {S.}~\bibnamefont {H\"{o}fling}}, \bibinfo {author}
  {\bibfnamefont {C.-Y.}\ \bibnamefont {Lu}},\ and\ \bibinfo {author}
  {\bibfnamefont {J.-W.}\ \bibnamefont {Pan}},\ }\bibfield  {title} {\bibinfo
  {title} {Coherently driving a single quantum two-level system with
  dichromatic laser pulses},\ }\href
  {https://doi.org/10.1038/s41567-019-0585-6} {\bibfield  {journal} {\bibinfo
  {journal} {Nature Physics}\ }\textbf {\bibinfo {volume} {15}},\ \bibinfo
  {pages} {941} (\bibinfo {year} {2019})}\BibitemShut {NoStop}%
\bibitem [{\citenamefont {Laferrière}\ \emph {et~al.}(2022)\citenamefont
  {Laferrière}, \citenamefont {Yeung}, \citenamefont {Miron}, \citenamefont
  {Northeast}, \citenamefont {Haffouz}, \citenamefont {Lapointe}, \citenamefont
  {Korkusinski}, \citenamefont {Poole}, \citenamefont {Williams},\ and\
  \citenamefont {Dalacu}}]{Laferriere2022}%
  \BibitemOpen
  \bibfield  {author} {\bibinfo {author} {\bibfnamefont {P.}~\bibnamefont
  {Laferrière}}, \bibinfo {author} {\bibfnamefont {E.}~\bibnamefont {Yeung}},
  \bibinfo {author} {\bibfnamefont {I.}~\bibnamefont {Miron}}, \bibinfo
  {author} {\bibfnamefont {D.~B.}\ \bibnamefont {Northeast}}, \bibinfo {author}
  {\bibfnamefont {S.}~\bibnamefont {Haffouz}}, \bibinfo {author} {\bibfnamefont
  {J.}~\bibnamefont {Lapointe}}, \bibinfo {author} {\bibfnamefont
  {M.}~\bibnamefont {Korkusinski}}, \bibinfo {author} {\bibfnamefont {P.~J.}\
  \bibnamefont {Poole}}, \bibinfo {author} {\bibfnamefont {R.~L.}\ \bibnamefont
  {Williams}},\ and\ \bibinfo {author} {\bibfnamefont {D.}~\bibnamefont
  {Dalacu}},\ }\bibfield  {title} {\bibinfo {title} {Unity yield of
  deterministically positioned quantum dot single photon sources},\ }\href
  {https://doi.org/10.1038/s41598-022-10451-1} {\bibfield  {journal} {\bibinfo
  {journal} {Scientific Reports}\ }\textbf {\bibinfo {volume} {12}},\ \bibinfo
  {pages} {6376} (\bibinfo {year} {2022})}\BibitemShut {NoStop}%
\bibitem [{\citenamefont {Appel}\ \emph {et~al.}(2022)\citenamefont {Appel},
  \citenamefont {Tiranov}, \citenamefont {Pabst}, \citenamefont {Chan},
  \citenamefont {Starup}, \citenamefont {Wang}, \citenamefont {Midolo},
  \citenamefont {Tiurev}, \citenamefont {Scholz}, \citenamefont {Wieck},
  \citenamefont {Ludwig}, \citenamefont {Sørensen},\ and\ \citenamefont
  {Lodahl}}]{Appel2022}%
  \BibitemOpen
  \bibfield  {author} {\bibinfo {author} {\bibfnamefont {M.~H.}\ \bibnamefont
  {Appel}}, \bibinfo {author} {\bibfnamefont {A.}~\bibnamefont {Tiranov}},
  \bibinfo {author} {\bibfnamefont {S.}~\bibnamefont {Pabst}}, \bibinfo
  {author} {\bibfnamefont {M.~L.}\ \bibnamefont {Chan}}, \bibinfo {author}
  {\bibfnamefont {C.}~\bibnamefont {Starup}}, \bibinfo {author} {\bibfnamefont
  {Y.}~\bibnamefont {Wang}}, \bibinfo {author} {\bibfnamefont {L.}~\bibnamefont
  {Midolo}}, \bibinfo {author} {\bibfnamefont {K.}~\bibnamefont {Tiurev}},
  \bibinfo {author} {\bibfnamefont {S.}~\bibnamefont {Scholz}}, \bibinfo
  {author} {\bibfnamefont {A.~D.}\ \bibnamefont {Wieck}}, \bibinfo {author}
  {\bibfnamefont {A.}~\bibnamefont {Ludwig}}, \bibinfo {author} {\bibfnamefont
  {A.~S.}\ \bibnamefont {Sørensen}},\ and\ \bibinfo {author} {\bibfnamefont
  {P.}~\bibnamefont {Lodahl}},\ }\bibfield  {title} {\bibinfo {title}
  {Entangling a {Hole} {Spin} with a {Time}-{Bin} {Photon}: {A} {Waveguide}
  {Approach} for {Quantum} {Dot} {Sources} of {Multiphoton} {Entanglement}},\
  }\href {https://doi.org/10.1103/PhysRevLett.128.233602} {\bibfield  {journal}
  {\bibinfo  {journal} {Physical Review Letters}\ }\textbf {\bibinfo {volume}
  {128}},\ \bibinfo {pages} {233602} (\bibinfo {year} {2022})}\BibitemShut
  {NoStop}%
\bibitem [{\citenamefont {Ginés}\ \emph {et~al.}(2022)\citenamefont {Ginés},
  \citenamefont {Moczała-Dusanowska}, \citenamefont {Dlaka}, \citenamefont
  {Hošák}, \citenamefont {Gonzales-Ureta}, \citenamefont {Lee}, \citenamefont
  {Ježek}, \citenamefont {Harbord}, \citenamefont {Oulton}, \citenamefont
  {Höfling}, \citenamefont {Young}, \citenamefont {Schneider},\ and\
  \citenamefont {Predojević}}]{Gines2022}%
  \BibitemOpen
  \bibfield  {author} {\bibinfo {author} {\bibfnamefont {L.}~\bibnamefont
  {Ginés}}, \bibinfo {author} {\bibfnamefont {M.}~\bibnamefont
  {Moczała-Dusanowska}}, \bibinfo {author} {\bibfnamefont {D.}~\bibnamefont
  {Dlaka}}, \bibinfo {author} {\bibfnamefont {R.}~\bibnamefont {Hošák}},
  \bibinfo {author} {\bibfnamefont {J.~R.}\ \bibnamefont {Gonzales-Ureta}},
  \bibinfo {author} {\bibfnamefont {J.}~\bibnamefont {Lee}}, \bibinfo {author}
  {\bibfnamefont {M.}~\bibnamefont {Ježek}}, \bibinfo {author} {\bibfnamefont
  {E.}~\bibnamefont {Harbord}}, \bibinfo {author} {\bibfnamefont
  {R.}~\bibnamefont {Oulton}}, \bibinfo {author} {\bibfnamefont
  {S.}~\bibnamefont {Höfling}}, \bibinfo {author} {\bibfnamefont {A.~B.}\
  \bibnamefont {Young}}, \bibinfo {author} {\bibfnamefont {C.}~\bibnamefont
  {Schneider}},\ and\ \bibinfo {author} {\bibfnamefont {A.}~\bibnamefont
  {Predojević}},\ }\bibfield  {title} {\bibinfo {title} {High {Extraction}
  {Efficiency} {Source} of {Photon} {Pairs} {Based} on a {Quantum} {Dot}
  {Embedded} in a {Broadband} {Micropillar} {Cavity}},\ }\href
  {https://doi.org/10.1103/PhysRevLett.129.033601} {\bibfield  {journal}
  {\bibinfo  {journal} {Physical Review Letters}\ }\textbf {\bibinfo {volume}
  {129}},\ \bibinfo {pages} {033601} (\bibinfo {year} {2022})}\BibitemShut
  {NoStop}%
\bibitem [{\citenamefont {Da~Lio}\ \emph {et~al.}(2022)\citenamefont {Da~Lio},
  \citenamefont {Faurby}, \citenamefont {Zhou}, \citenamefont {Chan},
  \citenamefont {Uppu}, \citenamefont {Thyrrestrup}, \citenamefont {Scholz},
  \citenamefont {Wieck}, \citenamefont {Ludwig}, \citenamefont {Lodahl},\ and\
  \citenamefont {Midolo}}]{Da_Lio2022}%
  \BibitemOpen
  \bibfield  {author} {\bibinfo {author} {\bibfnamefont {B.}~\bibnamefont
  {Da~Lio}}, \bibinfo {author} {\bibfnamefont {C.}~\bibnamefont {Faurby}},
  \bibinfo {author} {\bibfnamefont {X.}~\bibnamefont {Zhou}}, \bibinfo {author}
  {\bibfnamefont {M.~L.}\ \bibnamefont {Chan}}, \bibinfo {author}
  {\bibfnamefont {R.}~\bibnamefont {Uppu}}, \bibinfo {author} {\bibfnamefont
  {H.}~\bibnamefont {Thyrrestrup}}, \bibinfo {author} {\bibfnamefont
  {S.}~\bibnamefont {Scholz}}, \bibinfo {author} {\bibfnamefont {A.~D.}\
  \bibnamefont {Wieck}}, \bibinfo {author} {\bibfnamefont {A.}~\bibnamefont
  {Ludwig}}, \bibinfo {author} {\bibfnamefont {P.}~\bibnamefont {Lodahl}},\
  and\ \bibinfo {author} {\bibfnamefont {L.}~\bibnamefont {Midolo}},\
  }\bibfield  {title} {\bibinfo {title} {A {Pure} and {Indistinguishable}
  {Single}-{Photon} {Source} at {Telecommunication} {Wavelength}},\ }\href
  {https://doi.org/10.1002/qute.202200006} {\bibfield  {journal} {\bibinfo
  {journal} {Advanced Quantum Technologies}\ }\textbf {\bibinfo {volume} {5}},\
  \bibinfo {pages} {2200006} (\bibinfo {year} {2022})}\BibitemShut {NoStop}%
\bibitem [{\citenamefont {Bracht}\ \emph {et~al.}(2021)\citenamefont {Bracht},
  \citenamefont {Cosacchi}, \citenamefont {Seidelmann}, \citenamefont
  {Cygorek}, \citenamefont {Vagov}, \citenamefont {Axt}, \citenamefont
  {Heindel},\ and\ \citenamefont {Reiter}}]{Bracht2021}%
  \BibitemOpen
  \bibfield  {author} {\bibinfo {author} {\bibfnamefont {T.~K.}\ \bibnamefont
  {Bracht}}, \bibinfo {author} {\bibfnamefont {M.}~\bibnamefont {Cosacchi}},
  \bibinfo {author} {\bibfnamefont {T.}~\bibnamefont {Seidelmann}}, \bibinfo
  {author} {\bibfnamefont {M.}~\bibnamefont {Cygorek}}, \bibinfo {author}
  {\bibfnamefont {A.}~\bibnamefont {Vagov}}, \bibinfo {author} {\bibfnamefont
  {V.~M.}\ \bibnamefont {Axt}}, \bibinfo {author} {\bibfnamefont
  {T.}~\bibnamefont {Heindel}},\ and\ \bibinfo {author} {\bibfnamefont {D.~E.}\
  \bibnamefont {Reiter}},\ }\bibfield  {title} {\bibinfo {title} {Swing-{Up} of
  {Quantum} {Emitter} {Population} {Using} {Detuned} {Pulses}},\ }\href
  {https://doi.org/10.1103/PRXQuantum.2.040354} {\bibfield  {journal} {\bibinfo
   {journal} {PRX Quantum}\ }\textbf {\bibinfo {volume} {2}},\ \bibinfo {pages}
  {040354} (\bibinfo {year} {2021})}\BibitemShut {NoStop}%
\bibitem [{\citenamefont {Fischer}\ \emph {et~al.}(2016)\citenamefont
  {Fischer}, \citenamefont {Müller}, \citenamefont {Lagoudakis},\ and\
  \citenamefont {Vučković}}]{Fischer2016}%
  \BibitemOpen
  \bibfield  {author} {\bibinfo {author} {\bibfnamefont {K.~A.}\ \bibnamefont
  {Fischer}}, \bibinfo {author} {\bibfnamefont {K.}~\bibnamefont {Müller}},
  \bibinfo {author} {\bibfnamefont {K.~G.}\ \bibnamefont {Lagoudakis}},\ and\
  \bibinfo {author} {\bibfnamefont {J.}~\bibnamefont {Vučković}},\ }\bibfield
   {title} {\bibinfo {title} {Dynamical modeling of pulsed two-photon
  interference},\ }\href {https://doi.org/10.1088/1367-2630/18/11/113053}
  {\bibfield  {journal} {\bibinfo  {journal} {New Journal of Physics}\ }\textbf
  {\bibinfo {volume} {18}},\ \bibinfo {pages} {113053} (\bibinfo {year}
  {2016})}\BibitemShut {NoStop}%
\bibitem [{\citenamefont {Iles-Smith}\ \emph {et~al.}(2017)\citenamefont
  {Iles-Smith}, \citenamefont {McCutcheon}, \citenamefont {Nazir},\ and\
  \citenamefont {Mørk}}]{Iles-Smith2017}%
  \BibitemOpen
  \bibfield  {author} {\bibinfo {author} {\bibfnamefont {J.}~\bibnamefont
  {Iles-Smith}}, \bibinfo {author} {\bibfnamefont {D.~P.~S.}\ \bibnamefont
  {McCutcheon}}, \bibinfo {author} {\bibfnamefont {A.}~\bibnamefont {Nazir}},\
  and\ \bibinfo {author} {\bibfnamefont {J.}~\bibnamefont {Mørk}},\ }\bibfield
   {title} {\bibinfo {title} {Phonon scattering inhibits simultaneous
  near-unity efficiency and indistinguishability in semiconductor single-photon
  sources},\ }\href {https://doi.org/10.1038/nphoton.2017.101} {\bibfield
  {journal} {\bibinfo  {journal} {Nature Photonics}\ }\textbf {\bibinfo
  {volume} {11}},\ \bibinfo {pages} {521} (\bibinfo {year} {2017})}\BibitemShut
  {NoStop}%
\bibitem [{\citenamefont {Gustin}\ and\ \citenamefont
  {Hughes}(2018)}]{Gustin2018b}%
  \BibitemOpen
  \bibfield  {author} {\bibinfo {author} {\bibfnamefont {C.}~\bibnamefont
  {Gustin}}\ and\ \bibinfo {author} {\bibfnamefont {S.}~\bibnamefont
  {Hughes}},\ }\bibfield  {title} {\bibinfo {title} {Pulsed excitation dynamics
  in quantum-dot--cavity systems: Limits to optimizing the fidelity of
  on-demand single-photon sources},\ }\href
  {https://doi.org/10.1103/PhysRevB.98.045309} {\bibfield  {journal} {\bibinfo
  {journal} {Phys. Rev. B}\ }\textbf {\bibinfo {volume} {98}},\ \bibinfo
  {pages} {045309} (\bibinfo {year} {2018})}\BibitemShut {NoStop}%
\bibitem [{\citenamefont {Dorner}\ and\ \citenamefont
  {Zoller}(2002)}]{Dorner2002}%
  \BibitemOpen
  \bibfield  {author} {\bibinfo {author} {\bibfnamefont {U.}~\bibnamefont
  {Dorner}}\ and\ \bibinfo {author} {\bibfnamefont {P.}~\bibnamefont
  {Zoller}},\ }\bibfield  {title} {\bibinfo {title} {Laser-driven atoms in
  half-cavities},\ }\href {https://doi.org/10.1103/PhysRevA.66.023816}
  {\bibfield  {journal} {\bibinfo  {journal} {Physical Review A}\ }\textbf
  {\bibinfo {volume} {66}},\ \bibinfo {pages} {023816} (\bibinfo {year}
  {2002})}\BibitemShut {NoStop}%
\bibitem [{\citenamefont {Tufarelli}\ \emph {et~al.}(2013)\citenamefont
  {Tufarelli}, \citenamefont {Ciccarello},\ and\ \citenamefont
  {Kim}}]{Tufarelli2013}%
  \BibitemOpen
  \bibfield  {author} {\bibinfo {author} {\bibfnamefont {T.}~\bibnamefont
  {Tufarelli}}, \bibinfo {author} {\bibfnamefont {F.}~\bibnamefont
  {Ciccarello}},\ and\ \bibinfo {author} {\bibfnamefont {M.~S.}\ \bibnamefont
  {Kim}},\ }\bibfield  {title} {\bibinfo {title} {Dynamics of spontaneous
  emission in a single-end photonic waveguide},\ }\href
  {https://doi.org/10.1103/PhysRevA.87.013820} {\bibfield  {journal} {\bibinfo
  {journal} {Physical Review A}\ }\textbf {\bibinfo {volume} {87}},\ \bibinfo
  {pages} {013820} (\bibinfo {year} {2013})}\BibitemShut {NoStop}%
\bibitem [{\citenamefont {Carmele}\ \emph {et~al.}(2013)\citenamefont
  {Carmele}, \citenamefont {Kabuss}, \citenamefont {Schulze}, \citenamefont
  {Reitzenstein},\ and\ \citenamefont {Knorr}}]{Carmele2013}%
  \BibitemOpen
  \bibfield  {author} {\bibinfo {author} {\bibfnamefont {A.}~\bibnamefont
  {Carmele}}, \bibinfo {author} {\bibfnamefont {J.}~\bibnamefont {Kabuss}},
  \bibinfo {author} {\bibfnamefont {F.}~\bibnamefont {Schulze}}, \bibinfo
  {author} {\bibfnamefont {S.}~\bibnamefont {Reitzenstein}},\ and\ \bibinfo
  {author} {\bibfnamefont {A.}~\bibnamefont {Knorr}},\ }\bibfield  {title}
  {\bibinfo {title} {Single {Photon} {Delayed} {Feedback}: {A} {Way} to
  {Stabilize} {Intrinsic} {Quantum} {Cavity} {Electrodynamics}},\ }\href
  {https://doi.org/10.1103/PhysRevLett.110.013601} {\bibfield  {journal}
  {\bibinfo  {journal} {Physical Review Letters}\ }\textbf {\bibinfo {volume}
  {110}},\ \bibinfo {pages} {013601} (\bibinfo {year} {2013})}\BibitemShut
  {NoStop}%
\bibitem [{\citenamefont {Naumann}\ \emph {et~al.}(2016)\citenamefont
  {Naumann}, \citenamefont {Droenner}, \citenamefont {Hein}, \citenamefont
  {Carmele}, \citenamefont {Knorr},\ and\ \citenamefont
  {Kabuss}}]{Naumann2016}%
  \BibitemOpen
  \bibfield  {author} {\bibinfo {author} {\bibfnamefont {N.~L.}\ \bibnamefont
  {Naumann}}, \bibinfo {author} {\bibfnamefont {L.}~\bibnamefont {Droenner}},
  \bibinfo {author} {\bibfnamefont {S.~M.}\ \bibnamefont {Hein}}, \bibinfo
  {author} {\bibfnamefont {A.}~\bibnamefont {Carmele}}, \bibinfo {author}
  {\bibfnamefont {A.}~\bibnamefont {Knorr}},\ and\ \bibinfo {author}
  {\bibfnamefont {J.}~\bibnamefont {Kabuss}},\ }\bibfield  {title} {\bibinfo
  {title} {Feedback control of optomechanical systems},\ }in\ \href
  {https://doi.org/10.1117/12.2209338} {\emph {\bibinfo {booktitle} {Physics
  and {Simulation} of {Optoelectronic} {Devices} {XXIV}}}},\ Vol.\ \bibinfo
  {volume} {9742}\ (\bibinfo  {publisher} {International Society for Optics and
  Photonics},\ \bibinfo {year} {2016})\ p.\ \bibinfo {pages}
  {974216}\BibitemShut {NoStop}%
\bibitem [{\citenamefont {Lu}\ \emph {et~al.}(2017)\citenamefont {Lu},
  \citenamefont {Naumann}, \citenamefont {Cerrillo}, \citenamefont {Zhao},
  \citenamefont {Knorr},\ and\ \citenamefont {Carmele}}]{Lu2017}%
  \BibitemOpen
  \bibfield  {author} {\bibinfo {author} {\bibfnamefont {Y.}~\bibnamefont
  {Lu}}, \bibinfo {author} {\bibfnamefont {N.~L.}\ \bibnamefont {Naumann}},
  \bibinfo {author} {\bibfnamefont {J.}~\bibnamefont {Cerrillo}}, \bibinfo
  {author} {\bibfnamefont {Q.}~\bibnamefont {Zhao}}, \bibinfo {author}
  {\bibfnamefont {A.}~\bibnamefont {Knorr}},\ and\ \bibinfo {author}
  {\bibfnamefont {A.}~\bibnamefont {Carmele}},\ }\bibfield  {title} {\bibinfo
  {title} {Intensified antibunching via feedback-induced quantum
  interference},\ }\href {https://doi.org/10.1103/PhysRevA.95.063840}
  {\bibfield  {journal} {\bibinfo  {journal} {Physical Review A}\ }\textbf
  {\bibinfo {volume} {95}},\ \bibinfo {pages} {063840} (\bibinfo {year}
  {2017})}\BibitemShut {NoStop}%
\bibitem [{\citenamefont {Pichler}\ \emph {et~al.}(2017)\citenamefont
  {Pichler}, \citenamefont {Choi}, \citenamefont {Zoller},\ and\ \citenamefont
  {Lukin}}]{Pichler2017}%
  \BibitemOpen
  \bibfield  {author} {\bibinfo {author} {\bibfnamefont {H.}~\bibnamefont
  {Pichler}}, \bibinfo {author} {\bibfnamefont {S.}~\bibnamefont {Choi}},
  \bibinfo {author} {\bibfnamefont {P.}~\bibnamefont {Zoller}},\ and\ \bibinfo
  {author} {\bibfnamefont {M.~D.}\ \bibnamefont {Lukin}},\ }\bibfield  {title}
  {\bibinfo {title} {Universal photonic quantum computation via time-delayed
  feedback},\ }\href {https://doi.org/10.1073/pnas.1711003114} {\bibfield
  {journal} {\bibinfo  {journal} {Proceedings of the National Academy of
  Sciences}\ }\textbf {\bibinfo {volume} {114}},\ \bibinfo {pages} {11362}
  (\bibinfo {year} {2017})}\BibitemShut {NoStop}%
\bibitem [{\citenamefont {Wiseman}\ and\ \citenamefont
  {Milburn}(2002)}]{Wiseman2006}%
  \BibitemOpen
  \bibfield  {author} {\bibinfo {author} {\bibfnamefont {H.~M.}\ \bibnamefont
  {Wiseman}}\ and\ \bibinfo {author} {\bibfnamefont {G.~J.}\ \bibnamefont
  {Milburn}},\ }\href@noop {} {\emph {\bibinfo {title} {Quantum {M}easurement
  and {C}ontrol}}}\ (\bibinfo  {publisher} {Cambridge University Press},\
  \bibinfo {address} {Oxford},\ \bibinfo {year} {2002})\ p.\ \bibinfo {pages}
  {231}\BibitemShut {NoStop}%
\bibitem [{\citenamefont {Mu\~{n}oz Arias}\ \emph {et~al.}(2020)\citenamefont
  {Mu\~{n}oz Arias}, \citenamefont {Deutsch}, \citenamefont {Jessen},\ and\
  \citenamefont {Poggi}}]{Arias2020}%
  \BibitemOpen
  \bibfield  {author} {\bibinfo {author} {\bibfnamefont {M.~H.}\ \bibnamefont
  {Mu\~{n}oz Arias}}, \bibinfo {author} {\bibfnamefont {I.~H.}\ \bibnamefont
  {Deutsch}}, \bibinfo {author} {\bibfnamefont {P.~S.}\ \bibnamefont
  {Jessen}},\ and\ \bibinfo {author} {\bibfnamefont {P.~M.}\ \bibnamefont
  {Poggi}},\ }\bibfield  {title} {\bibinfo {title} {Simulation of complex
  dynamics of mean-field $p$-spin models using measurement-based quantum
  feedback control},\ }\href {https://doi.org/10.1103/PhysRevA.102.022610}
  {\bibfield  {journal} {\bibinfo  {journal} {Physical Review A}\ }\textbf
  {\bibinfo {volume} {102}},\ \bibinfo {pages} {022610} (\bibinfo {year}
  {2020})}\BibitemShut {NoStop}%
\bibitem [{\citenamefont {Grigoletto}\ and\ \citenamefont
  {Ticozzi}(2021)}]{Grigoletto2021}%
  \BibitemOpen
  \bibfield  {author} {\bibinfo {author} {\bibfnamefont {T.}~\bibnamefont
  {Grigoletto}}\ and\ \bibinfo {author} {\bibfnamefont {F.}~\bibnamefont
  {Ticozzi}},\ }\bibfield  {title} {\bibinfo {title} {Stabilization {Via}
  {Feedback} {Switching} for {Quantum} {Stochastic} {Dynamics}},\ }\href
  {https://doi.org/10.1109/LCSYS.2021.3065603} {\bibfield  {journal} {\bibinfo
  {journal} {IEEE Control Systems Letters}\ }\textbf {\bibinfo {volume} {6}},\
  \bibinfo {pages} {235} (\bibinfo {year} {2021})}\BibitemShut {NoStop}%
\bibitem [{\citenamefont {{Hjelme}}\ \emph {et~al.}(1991)\citenamefont
  {{Hjelme}}, \citenamefont {{Mickelson}},\ and\ \citenamefont
  {{Beausoleil}}}]{Hjelme1991}%
  \BibitemOpen
  \bibfield  {author} {\bibinfo {author} {\bibfnamefont {D.~R.}\ \bibnamefont
  {{Hjelme}}}, \bibinfo {author} {\bibfnamefont {A.~R.}\ \bibnamefont
  {{Mickelson}}},\ and\ \bibinfo {author} {\bibfnamefont {R.~G.}\ \bibnamefont
  {{Beausoleil}}},\ }\bibfield  {title} {\bibinfo {title} {Semiconductor laser
  stabilization by external optical feedback},\ }\href
  {https://ieeexplore.ieee.org/document/81333} {\bibfield  {journal} {\bibinfo
  {journal} {IEEE Journal of Quantum Electronics}\ }\textbf {\bibinfo {volume}
  {27}},\ \bibinfo {pages} {352} (\bibinfo {year} {1991})}\BibitemShut
  {NoStop}%
\bibitem [{\citenamefont {Franklin}\ \emph {et~al.}(2014)\citenamefont
  {Franklin}, \citenamefont {Powell},\ and\ \citenamefont
  {Emami-Naeini}}]{Franklin2014}%
  \BibitemOpen
  \bibfield  {author} {\bibinfo {author} {\bibfnamefont {G.~F.}\ \bibnamefont
  {Franklin}}, \bibinfo {author} {\bibfnamefont {J.~D.}\ \bibnamefont
  {Powell}},\ and\ \bibinfo {author} {\bibfnamefont {A.}~\bibnamefont
  {Emami-Naeini}},\ }\href@noop {} {\emph {\bibinfo {title} {Feedback Control
  of Dynamic Systems}}}\ (\bibinfo  {publisher} {Pearson},\ \bibinfo {year}
  {2014})\BibitemShut {NoStop}%
\bibitem [{\citenamefont {Kubanek}\ \emph {et~al.}(2009)\citenamefont
  {Kubanek}, \citenamefont {Koch}, \citenamefont {Sames}, \citenamefont
  {Ourjoumtsev}, \citenamefont {Pinkse}, \citenamefont {Murr},\ and\
  \citenamefont {Rempe}}]{Kubanek2009}%
  \BibitemOpen
  \bibfield  {author} {\bibinfo {author} {\bibfnamefont {A.}~\bibnamefont
  {Kubanek}}, \bibinfo {author} {\bibfnamefont {M.}~\bibnamefont {Koch}},
  \bibinfo {author} {\bibfnamefont {C.}~\bibnamefont {Sames}}, \bibinfo
  {author} {\bibfnamefont {A.}~\bibnamefont {Ourjoumtsev}}, \bibinfo {author}
  {\bibfnamefont {P.~W.~H.}\ \bibnamefont {Pinkse}}, \bibinfo {author}
  {\bibfnamefont {K.}~\bibnamefont {Murr}},\ and\ \bibinfo {author}
  {\bibfnamefont {G.}~\bibnamefont {Rempe}},\ }\bibfield  {title} {\bibinfo
  {title} {Photon-by-photon feedback control of a single-atom trajectory},\
  }\href {https://doi.org/10.1038/nature08563} {\bibfield  {journal} {\bibinfo
  {journal} {Nature}\ }\textbf {\bibinfo {volume} {462}},\ \bibinfo {pages}
  {898} (\bibinfo {year} {2009})}\BibitemShut {NoStop}%
\bibitem [{\citenamefont {Gillett}\ \emph {et~al.}(2010)\citenamefont
  {Gillett}, \citenamefont {Dalton}, \citenamefont {Lanyon}, \citenamefont
  {Almeida}, \citenamefont {Barbieri}, \citenamefont {Pryde}, \citenamefont
  {O'Brien}, \citenamefont {Resch}, \citenamefont {Bartlett},\ and\
  \citenamefont {White}}]{Gillett2010}%
  \BibitemOpen
  \bibfield  {author} {\bibinfo {author} {\bibfnamefont {G.~G.}\ \bibnamefont
  {Gillett}}, \bibinfo {author} {\bibfnamefont {R.~B.}\ \bibnamefont {Dalton}},
  \bibinfo {author} {\bibfnamefont {B.~P.}\ \bibnamefont {Lanyon}}, \bibinfo
  {author} {\bibfnamefont {M.~P.}\ \bibnamefont {Almeida}}, \bibinfo {author}
  {\bibfnamefont {M.}~\bibnamefont {Barbieri}}, \bibinfo {author}
  {\bibfnamefont {G.~J.}\ \bibnamefont {Pryde}}, \bibinfo {author}
  {\bibfnamefont {J.~L.}\ \bibnamefont {O'Brien}}, \bibinfo {author}
  {\bibfnamefont {K.~J.}\ \bibnamefont {Resch}}, \bibinfo {author}
  {\bibfnamefont {S.~D.}\ \bibnamefont {Bartlett}},\ and\ \bibinfo {author}
  {\bibfnamefont {A.~G.}\ \bibnamefont {White}},\ }\bibfield  {title} {\bibinfo
  {title} {Experimental feedback control of quantum systems using weak
  measurements},\ }\href {https://doi.org/10.1103/PhysRevLett.104.080503}
  {\bibfield  {journal} {\bibinfo  {journal} {Phys. Rev. Lett.}\ }\textbf
  {\bibinfo {volume} {104}},\ \bibinfo {pages} {080503} (\bibinfo {year}
  {2010})}\BibitemShut {NoStop}%
\bibitem [{\citenamefont {Rafiee}\ \emph {et~al.}(2020)\citenamefont {Rafiee},
  \citenamefont {Nourmandipour},\ and\ \citenamefont {Mancini}}]{Rafiee2020}%
  \BibitemOpen
  \bibfield  {author} {\bibinfo {author} {\bibfnamefont {M.}~\bibnamefont
  {Rafiee}}, \bibinfo {author} {\bibfnamefont {A.}~\bibnamefont
  {Nourmandipour}},\ and\ \bibinfo {author} {\bibfnamefont {S.}~\bibnamefont
  {Mancini}},\ }\bibfield  {title} {\bibinfo {title} {Enforcing dissipative
  entanglement by feedback},\ }\href
  {https://doi.org/10.1016/j.physleta.2020.126748} {\bibfield  {journal}
  {\bibinfo  {journal} {Physics Letters A}\ }\textbf {\bibinfo {volume}
  {384}},\ \bibinfo {pages} {126748} (\bibinfo {year} {2020})}\BibitemShut
  {NoStop}%
\bibitem [{\citenamefont {Tan}\ \emph {et~al.}(2021)\citenamefont {Tan},
  \citenamefont {Camati}, \citenamefont {Cauquil}, \citenamefont {Auff\`eves},\
  and\ \citenamefont {Dotsenko}}]{Tan2020}%
  \BibitemOpen
  \bibfield  {author} {\bibinfo {author} {\bibfnamefont {Z.}~\bibnamefont
  {Tan}}, \bibinfo {author} {\bibfnamefont {P.~A.}\ \bibnamefont {Camati}},
  \bibinfo {author} {\bibfnamefont {G.~C.}\ \bibnamefont {Cauquil}}, \bibinfo
  {author} {\bibfnamefont {A.}~\bibnamefont {Auff\`eves}},\ and\ \bibinfo
  {author} {\bibfnamefont {I.}~\bibnamefont {Dotsenko}},\ }\bibfield  {title}
  {\bibinfo {title} {Alternative experimental ways to access entropy
  production},\ }\href {https://doi.org/10.1103/PhysRevResearch.3.043076}
  {\bibfield  {journal} {\bibinfo  {journal} {Phys. Rev. Research}\ }\textbf
  {\bibinfo {volume} {3}},\ \bibinfo {pages} {043076} (\bibinfo {year}
  {2021})}\BibitemShut {NoStop}%
\bibitem [{\citenamefont {Di~Giovanni}\ \emph {et~al.}(2021)\citenamefont
  {Di~Giovanni}, \citenamefont {Brunelli},\ and\ \citenamefont
  {Genoni}}]{Giovanni2021}%
  \BibitemOpen
  \bibfield  {author} {\bibinfo {author} {\bibfnamefont {A.}~\bibnamefont
  {Di~Giovanni}}, \bibinfo {author} {\bibfnamefont {M.}~\bibnamefont
  {Brunelli}},\ and\ \bibinfo {author} {\bibfnamefont {M.~G.}\ \bibnamefont
  {Genoni}},\ }\bibfield  {title} {\bibinfo {title} {Unconditional mechanical
  squeezing via backaction-evading measurements and nonoptimal feedback
  control},\ }\href {https://doi.org/10.1103/PhysRevA.103.022614} {\bibfield
  {journal} {\bibinfo  {journal} {Physical Review A}\ }\textbf {\bibinfo
  {volume} {103}},\ \bibinfo {pages} {022614} (\bibinfo {year}
  {2021})}\BibitemShut {NoStop}%
\bibitem [{\citenamefont {Koshino}\ and\ \citenamefont
  {Nakamura}(2012)}]{Koshino2012}%
  \BibitemOpen
  \bibfield  {author} {\bibinfo {author} {\bibfnamefont {K.}~\bibnamefont
  {Koshino}}\ and\ \bibinfo {author} {\bibfnamefont {Y.}~\bibnamefont
  {Nakamura}},\ }\bibfield  {title} {\bibinfo {title} {Control of the radiative
  level shift and linewidth of a superconducting artificial atom through a
  variable boundary condition},\ }\href
  {https://doi.org/10.1088/1367-2630/14/4/043005} {\bibfield  {journal}
  {\bibinfo  {journal} {New Journal of Physics}\ }\textbf {\bibinfo {volume}
  {14}},\ \bibinfo {pages} {043005} (\bibinfo {year} {2012})}\BibitemShut
  {NoStop}%
\bibitem [{\citenamefont {Hoi}\ \emph {et~al.}(2015)\citenamefont {Hoi},
  \citenamefont {Kockum}, \citenamefont {Tornberg}, \citenamefont
  {Pourkabirian}, \citenamefont {Johansson}, \citenamefont {Delsing},\ and\
  \citenamefont {Wilson}}]{Hoi2015}%
  \BibitemOpen
  \bibfield  {author} {\bibinfo {author} {\bibfnamefont {I.-C.}\ \bibnamefont
  {Hoi}}, \bibinfo {author} {\bibfnamefont {A.~F.}\ \bibnamefont {Kockum}},
  \bibinfo {author} {\bibfnamefont {L.}~\bibnamefont {Tornberg}}, \bibinfo
  {author} {\bibfnamefont {A.}~\bibnamefont {Pourkabirian}}, \bibinfo {author}
  {\bibfnamefont {G.}~\bibnamefont {Johansson}}, \bibinfo {author}
  {\bibfnamefont {P.}~\bibnamefont {Delsing}},\ and\ \bibinfo {author}
  {\bibfnamefont {C.~M.}\ \bibnamefont {Wilson}},\ }\bibfield  {title}
  {\bibinfo {title} {Probing the quantum vacuum with an artificial atom in
  front of a mirror},\ }\href {https://doi.org/10.1038/nphys3484} {\bibfield
  {journal} {\bibinfo  {journal} {Nature Physics}\ }\textbf {\bibinfo {volume}
  {11}},\ \bibinfo {pages} {1045} (\bibinfo {year} {2015})}\BibitemShut
  {NoStop}%
\bibitem [{\citenamefont {Grimsmo}(2015)}]{Grimsmo2015}%
  \BibitemOpen
  \bibfield  {author} {\bibinfo {author} {\bibfnamefont {A.~L.}\ \bibnamefont
  {Grimsmo}},\ }\bibfield  {title} {\bibinfo {title} {Time-{Delayed} {Quantum}
  {Feedback} {Control}},\ }\href
  {https://doi.org/10.1103/PhysRevLett.115.060402} {\bibfield  {journal}
  {\bibinfo  {journal} {Physical Review Letters}\ }\textbf {\bibinfo {volume}
  {115}},\ \bibinfo {pages} {060402} (\bibinfo {year} {2015})}\BibitemShut
  {NoStop}%
\bibitem [{\citenamefont {Kabuss}\ \emph {et~al.}(2016)\citenamefont {Kabuss},
  \citenamefont {Katsch}, \citenamefont {Knorr},\ and\ \citenamefont
  {Carmele}}]{Kabuss2016}%
  \BibitemOpen
  \bibfield  {author} {\bibinfo {author} {\bibfnamefont {J.}~\bibnamefont
  {Kabuss}}, \bibinfo {author} {\bibfnamefont {F.}~\bibnamefont {Katsch}},
  \bibinfo {author} {\bibfnamefont {A.}~\bibnamefont {Knorr}},\ and\ \bibinfo
  {author} {\bibfnamefont {A.}~\bibnamefont {Carmele}},\ }\bibfield  {title}
  {\bibinfo {title} {Unraveling coherent quantum feedback for {Pyragas}
  control},\ }\href {https://doi.org/10.1364/JOSAB.33.000C10} {\bibfield
  {journal} {\bibinfo  {journal} {Journal of the Optical Society of America B}\
  }\textbf {\bibinfo {volume} {33}},\ \bibinfo {pages} {C10} (\bibinfo {year}
  {2016})}\BibitemShut {NoStop}%
\bibitem [{\citenamefont {Pichler}\ and\ \citenamefont
  {Zoller}(2016)}]{Pichler2016}%
  \BibitemOpen
  \bibfield  {author} {\bibinfo {author} {\bibfnamefont {H.}~\bibnamefont
  {Pichler}}\ and\ \bibinfo {author} {\bibfnamefont {P.}~\bibnamefont
  {Zoller}},\ }\bibfield  {title} {\bibinfo {title} {Photonic {Circuits} with
  {Time} {Delays} and {Quantum} {Feedback}},\ }\href
  {https://doi.org/10.1103/PhysRevLett.116.093601} {\bibfield  {journal}
  {\bibinfo  {journal} {Physical Review Letters}\ }\textbf {\bibinfo {volume}
  {116}},\ \bibinfo {pages} {093601} (\bibinfo {year} {2016})}\BibitemShut
  {NoStop}%
\bibitem [{\citenamefont {N\'{e}met}\ and\ \citenamefont
  {Parkins}(2016)}]{Nemet2016}%
  \BibitemOpen
  \bibfield  {author} {\bibinfo {author} {\bibfnamefont {N.}~\bibnamefont
  {N\'{e}met}}\ and\ \bibinfo {author} {\bibfnamefont {S.}~\bibnamefont
  {Parkins}},\ }\bibfield  {title} {\bibinfo {title} {Enhanced optical
  squeezing from a degenerate parametric amplifier via time-delayed coherent
  feedback},\ }\href {https://doi.org/10.1103/PhysRevA.94.023809} {\bibfield
  {journal} {\bibinfo  {journal} {Physical Review A}\ }\textbf {\bibinfo
  {volume} {94}},\ \bibinfo {pages} {023809} (\bibinfo {year}
  {2016})}\BibitemShut {NoStop}%
\bibitem [{\citenamefont {Hein}\ \emph {et~al.}(2016)\citenamefont {Hein},
  \citenamefont {Carmele},\ and\ \citenamefont {Knorr}}]{Hein2016}%
  \BibitemOpen
  \bibfield  {author} {\bibinfo {author} {\bibfnamefont {S.~M.}\ \bibnamefont
  {Hein}}, \bibinfo {author} {\bibfnamefont {A.}~\bibnamefont {Carmele}},\ and\
  \bibinfo {author} {\bibfnamefont {A.}~\bibnamefont {Knorr}},\ }\bibfield
  {title} {\bibinfo {title} {Creation and control of entanglement by
  time-delayed quantum-coherent feedback},\ }in\ \href
  {https://doi.org/10.1117/12.2207671} {\emph {\bibinfo {booktitle} {Physics
  and {Simulation} of {Optoelectronic} {Devices} {XXIV}}}},\ Vol.\ \bibinfo
  {volume} {9742}\ (\bibinfo  {publisher} {International Society for Optics and
  Photonics},\ \bibinfo {year} {2016})\ p.\ \bibinfo {pages}
  {97420X}\BibitemShut {NoStop}%
\bibitem [{\citenamefont {Guimond}\ \emph {et~al.}(2016)\citenamefont
  {Guimond}, \citenamefont {Pichler}, \citenamefont {Rauschenbeutel},\ and\
  \citenamefont {Zoller}}]{Guimond2016}%
  \BibitemOpen
  \bibfield  {author} {\bibinfo {author} {\bibfnamefont {P.-O.}\ \bibnamefont
  {Guimond}}, \bibinfo {author} {\bibfnamefont {H.}~\bibnamefont {Pichler}},
  \bibinfo {author} {\bibfnamefont {A.}~\bibnamefont {Rauschenbeutel}},\ and\
  \bibinfo {author} {\bibfnamefont {P.}~\bibnamefont {Zoller}},\ }\bibfield
  {title} {\bibinfo {title} {Chiral quantum optics with {V}-level atoms and
  coherent quantum feedback},\ }\href
  {https://doi.org/10.1103/PhysRevA.94.033829} {\bibfield  {journal} {\bibinfo
  {journal} {Physical Review A}\ }\textbf {\bibinfo {volume} {94}},\ \bibinfo
  {pages} {033829} (\bibinfo {year} {2016})}\BibitemShut {NoStop}%
\bibitem [{\citenamefont {Whalen}\ \emph {et~al.}(2017)\citenamefont {Whalen},
  \citenamefont {Grimsmo},\ and\ \citenamefont {Carmichael}}]{Whalen2017}%
  \BibitemOpen
  \bibfield  {author} {\bibinfo {author} {\bibfnamefont {S.~J.}\ \bibnamefont
  {Whalen}}, \bibinfo {author} {\bibfnamefont {A.~L.}\ \bibnamefont
  {Grimsmo}},\ and\ \bibinfo {author} {\bibfnamefont {H.~J.}\ \bibnamefont
  {Carmichael}},\ }\bibfield  {title} {\bibinfo {title} {Open quantum systems
  with delayed coherent feedback},\ }\href
  {https://doi.org/10.1088/2058-9565/aa8331} {\bibfield  {journal} {\bibinfo
  {journal} {Quantum Science and Technology}\ }\textbf {\bibinfo {volume}
  {2}},\ \bibinfo {pages} {044008} (\bibinfo {year} {2017})}\BibitemShut
  {NoStop}%
\bibitem [{\citenamefont {Naumann}\ \emph {et~al.}(2017)\citenamefont
  {Naumann}, \citenamefont {Hein}, \citenamefont {Kraft}, \citenamefont
  {Knorr},\ and\ \citenamefont {Carmele}}]{Naumann2017}%
  \BibitemOpen
  \bibfield  {author} {\bibinfo {author} {\bibfnamefont {N.~L.}\ \bibnamefont
  {Naumann}}, \bibinfo {author} {\bibfnamefont {S.~M.}\ \bibnamefont {Hein}},
  \bibinfo {author} {\bibfnamefont {M.}~\bibnamefont {Kraft}}, \bibinfo
  {author} {\bibfnamefont {A.}~\bibnamefont {Knorr}},\ and\ \bibinfo {author}
  {\bibfnamefont {A.}~\bibnamefont {Carmele}},\ }\bibfield  {title} {\bibinfo
  {title} {Feedback control of photon statistics},\ }in\ \href
  {https://doi.org/10.1117/12.2251952} {\emph {\bibinfo {booktitle} {Physics
  and {Simulation} of {Optoelectronic} {Devices} {XXV}}}},\ Vol.\ \bibinfo
  {volume} {10098}\ (\bibinfo  {publisher} {International Society for Optics
  and Photonics},\ \bibinfo {year} {2017})\ p.\ \bibinfo {pages}
  {100980N}\BibitemShut {NoStop}%
\bibitem [{\citenamefont {Guimond}\ \emph {et~al.}(2017)\citenamefont
  {Guimond}, \citenamefont {Pletyukhov}, \citenamefont {Pichler},\ and\
  \citenamefont {Zoller}}]{Guimond2017}%
  \BibitemOpen
  \bibfield  {author} {\bibinfo {author} {\bibfnamefont {P.-O.}\ \bibnamefont
  {Guimond}}, \bibinfo {author} {\bibfnamefont {M.}~\bibnamefont {Pletyukhov}},
  \bibinfo {author} {\bibfnamefont {H.}~\bibnamefont {Pichler}},\ and\ \bibinfo
  {author} {\bibfnamefont {P.}~\bibnamefont {Zoller}},\ }\bibfield  {title}
  {\bibinfo {title} {Delayed coherent quantum feedback from a scattering theory
  and a matrix product state perspective},\ }\href
  {https://doi.org/10.1088/2058-9565/aa7f03} {\bibfield  {journal} {\bibinfo
  {journal} {Quantum Science and Technology}\ }\textbf {\bibinfo {volume}
  {2}},\ \bibinfo {pages} {044012} (\bibinfo {year} {2017})}\BibitemShut
  {NoStop}%
\bibitem [{\citenamefont {Forn-Díaz}\ \emph {et~al.}(2017)\citenamefont
  {Forn-Díaz}, \citenamefont {Warren}, \citenamefont {Chang}, \citenamefont
  {Vadiraj},\ and\ \citenamefont {Wilson}}]{Forn2017}%
  \BibitemOpen
  \bibfield  {author} {\bibinfo {author} {\bibfnamefont {P.}~\bibnamefont
  {Forn-Díaz}}, \bibinfo {author} {\bibfnamefont {C.}~\bibnamefont {Warren}},
  \bibinfo {author} {\bibfnamefont {C.}~\bibnamefont {Chang}}, \bibinfo
  {author} {\bibfnamefont {A.}~\bibnamefont {Vadiraj}},\ and\ \bibinfo {author}
  {\bibfnamefont {C.}~\bibnamefont {Wilson}},\ }\bibfield  {title} {\bibinfo
  {title} {On-{Demand} {Microwave} {Generator} of {Shaped} {Single}
  {Photons}},\ }\href {https://doi.org/10.1103/PhysRevApplied.8.054015}
  {\bibfield  {journal} {\bibinfo  {journal} {Physical Review Applied}\
  }\textbf {\bibinfo {volume} {8}},\ \bibinfo {pages} {054015} (\bibinfo {year}
  {2017})}\BibitemShut {NoStop}%
\bibitem [{\citenamefont {Whalen}(2019)}]{Whalen2019}%
  \BibitemOpen
  \bibfield  {author} {\bibinfo {author} {\bibfnamefont {S.~J.}\ \bibnamefont
  {Whalen}},\ }\bibfield  {title} {\bibinfo {title} {Collision model for
  non-{Markovian} quantum trajectories},\ }\href
  {https://doi.org/10.1103/PhysRevA.100.052113} {\bibfield  {journal} {\bibinfo
   {journal} {Physical Review A}\ }\textbf {\bibinfo {volume} {100}},\ \bibinfo
  {pages} {052113} (\bibinfo {year} {2019})}\BibitemShut {NoStop}%
\bibitem [{\citenamefont {N\'{e}met}\ \emph {et~al.}(2019)\citenamefont
  {N\'{e}met}, \citenamefont {Parkins}, \citenamefont {Knorr},\ and\
  \citenamefont {Carmele}}]{Nemet2019}%
  \BibitemOpen
  \bibfield  {author} {\bibinfo {author} {\bibfnamefont {N.}~\bibnamefont
  {N\'{e}met}}, \bibinfo {author} {\bibfnamefont {S.}~\bibnamefont {Parkins}},
  \bibinfo {author} {\bibfnamefont {A.}~\bibnamefont {Knorr}},\ and\ \bibinfo
  {author} {\bibfnamefont {A.}~\bibnamefont {Carmele}},\ }\bibfield  {title}
  {\bibinfo {title} {Stabilizing quantum coherence against pure dephasing in
  the presence of time-delayed coherent feedback at finite temperature},\
  }\href {https://doi.org/10.1103/PhysRevA.99.053809} {\bibfield  {journal}
  {\bibinfo  {journal} {Physical Review A}\ }\textbf {\bibinfo {volume} {99}},\
  \bibinfo {pages} {053809} (\bibinfo {year} {2019})}\BibitemShut {NoStop}%
\bibitem [{\citenamefont {Calaj\'{o}}\ \emph {et~al.}(2019)\citenamefont
  {Calaj\'{o}}, \citenamefont {Fang}, \citenamefont {Baranger},\ and\
  \citenamefont {Ciccarello}}]{Calajo2019}%
  \BibitemOpen
  \bibfield  {author} {\bibinfo {author} {\bibfnamefont {G.}~\bibnamefont
  {Calaj\'{o}}}, \bibinfo {author} {\bibfnamefont {Y.-L.~L.}\ \bibnamefont
  {Fang}}, \bibinfo {author} {\bibfnamefont {H.~U.}\ \bibnamefont {Baranger}},\
  and\ \bibinfo {author} {\bibfnamefont {F.}~\bibnamefont {Ciccarello}},\
  }\bibfield  {title} {\bibinfo {title} {Exciting a {Bound} {State} in the
  {Continuum} through {Multiphoton} {Scattering} {Plus} {Delayed} {Quantum}
  {Feedback}},\ }\href {https://doi.org/10.1103/PhysRevLett.122.073601}
  {\bibfield  {journal} {\bibinfo  {journal} {Physical Review Letters}\
  }\textbf {\bibinfo {volume} {122}},\ \bibinfo {pages} {073601} (\bibinfo
  {year} {2019})}\BibitemShut {NoStop}%
\bibitem [{\citenamefont {Crowder}\ \emph {et~al.}(2020)\citenamefont
  {Crowder}, \citenamefont {Carmichael},\ and\ \citenamefont
  {Hughes}}]{Crowder2020}%
  \BibitemOpen
  \bibfield  {author} {\bibinfo {author} {\bibfnamefont {G.}~\bibnamefont
  {Crowder}}, \bibinfo {author} {\bibfnamefont {H.}~\bibnamefont
  {Carmichael}},\ and\ \bibinfo {author} {\bibfnamefont {S.}~\bibnamefont
  {Hughes}},\ }\bibfield  {title} {\bibinfo {title} {Quantum trajectory theory
  of few-photon cavity-{QED} systems with a time-delayed coherent feedback},\
  }\href {https://doi.org/10.1103/PhysRevA.101.023807} {\bibfield  {journal}
  {\bibinfo  {journal} {Physical Review A}\ }\textbf {\bibinfo {volume}
  {101}},\ \bibinfo {pages} {023807} (\bibinfo {year} {2020})}\BibitemShut
  {NoStop}%
\bibitem [{\citenamefont {Harwood}\ \emph {et~al.}(2021)\citenamefont
  {Harwood}, \citenamefont {Brunelli},\ and\ \citenamefont
  {Serafini}}]{Harwood2021}%
  \BibitemOpen
  \bibfield  {author} {\bibinfo {author} {\bibfnamefont {A.}~\bibnamefont
  {Harwood}}, \bibinfo {author} {\bibfnamefont {M.}~\bibnamefont {Brunelli}},\
  and\ \bibinfo {author} {\bibfnamefont {A.}~\bibnamefont {Serafini}},\
  }\bibfield  {title} {\bibinfo {title} {Cavity optomechanics assisted by
  optical coherent feedback},\ }\href
  {https://doi.org/10.1103/PhysRevA.103.023509} {\bibfield  {journal} {\bibinfo
   {journal} {Physical Review A}\ }\textbf {\bibinfo {volume} {103}},\ \bibinfo
  {pages} {023509} (\bibinfo {year} {2021})}\BibitemShut {NoStop}%
\bibitem [{\citenamefont {Barkemeyer}\ \emph {et~al.}(2021)\citenamefont
  {Barkemeyer}, \citenamefont {Hohn}, \citenamefont {Reitzenstein},\ and\
  \citenamefont {Carmele}}]{Barkemeyer2021}%
  \BibitemOpen
  \bibfield  {author} {\bibinfo {author} {\bibfnamefont {K.}~\bibnamefont
  {Barkemeyer}}, \bibinfo {author} {\bibfnamefont {M.}~\bibnamefont {Hohn}},
  \bibinfo {author} {\bibfnamefont {S.}~\bibnamefont {Reitzenstein}},\ and\
  \bibinfo {author} {\bibfnamefont {A.}~\bibnamefont {Carmele}},\ }\bibfield
  {title} {\bibinfo {title} {Boosting energy-time entanglement using coherent
  time-delayed feedback},\ }\href {https://doi.org/10.1103/PhysRevA.103.062423}
  {\bibfield  {journal} {\bibinfo  {journal} {Physical Review A}\ }\textbf
  {\bibinfo {volume} {103}},\ \bibinfo {pages} {062423} (\bibinfo {year}
  {2021})}\BibitemShut {NoStop}%
\bibitem [{\citenamefont {Shi}\ and\ \citenamefont {Waks}(2021)}]{Shi2021}%
  \BibitemOpen
  \bibfield  {author} {\bibinfo {author} {\bibfnamefont {Y.}~\bibnamefont
  {Shi}}\ and\ \bibinfo {author} {\bibfnamefont {E.}~\bibnamefont {Waks}},\
  }\bibfield  {title} {\bibinfo {title} {Deterministic generation of
  multidimensional photonic cluster states using time-delay feedback},\ }\href
  {https://doi.org/10.1103/PhysRevA.104.013703} {\bibfield  {journal} {\bibinfo
   {journal} {Physical Review A}\ }\textbf {\bibinfo {volume} {104}},\ \bibinfo
  {pages} {013703} (\bibinfo {year} {2021})}\BibitemShut {NoStop}%
\bibitem [{\citenamefont {Arranz~Regidor}\ \emph {et~al.}(2021)\citenamefont
  {Arranz~Regidor}, \citenamefont {Crowder}, \citenamefont {Carmichael},\ and\
  \citenamefont {Hughes}}]{Regidor2021a}%
  \BibitemOpen
  \bibfield  {author} {\bibinfo {author} {\bibfnamefont {S.}~\bibnamefont
  {Arranz~Regidor}}, \bibinfo {author} {\bibfnamefont {G.}~\bibnamefont
  {Crowder}}, \bibinfo {author} {\bibfnamefont {H.}~\bibnamefont
  {Carmichael}},\ and\ \bibinfo {author} {\bibfnamefont {S.}~\bibnamefont
  {Hughes}},\ }\bibfield  {title} {\bibinfo {title} {Modeling quantum
  light-matter interactions in waveguide {QED} with retardation, nonlinear
  interactions, and a time-delayed feedback: {Matrix} product states versus a
  space-discretized waveguide model},\ }\href
  {https://doi.org/10.1103/PhysRevResearch.3.023030} {\bibfield  {journal}
  {\bibinfo  {journal} {Physical Review Research}\ }\textbf {\bibinfo {volume}
  {3}},\ \bibinfo {pages} {023030} (\bibinfo {year} {2021})}\BibitemShut
  {NoStop}%
\bibitem [{\citenamefont {Barkemeyer}\ \emph {et~al.}(2022)\citenamefont
  {Barkemeyer}, \citenamefont {Knorr},\ and\ \citenamefont
  {Carmele}}]{Barkemeyer2022}%
  \BibitemOpen
  \bibfield  {author} {\bibinfo {author} {\bibfnamefont {K.}~\bibnamefont
  {Barkemeyer}}, \bibinfo {author} {\bibfnamefont {A.}~\bibnamefont {Knorr}},\
  and\ \bibinfo {author} {\bibfnamefont {A.}~\bibnamefont {Carmele}},\
  }\bibfield  {title} {\bibinfo {title} {Heisenberg treatment of multiphoton
  pulses in waveguide {QED} with time-delayed feedback},\ }\href
  {https://doi.org/10.1103/PhysRevA.106.023708} {\bibfield  {journal} {\bibinfo
   {journal} {Physical Review A}\ }\textbf {\bibinfo {volume} {106}},\ \bibinfo
  {pages} {023708} (\bibinfo {year} {2022})}\BibitemShut {NoStop}%
\bibitem [{\citenamefont {Crowder}\ \emph {et~al.}(2022)\citenamefont
  {Crowder}, \citenamefont {Ramunno},\ and\ \citenamefont
  {Hughes}}]{Crowder2022}%
  \BibitemOpen
  \bibfield  {author} {\bibinfo {author} {\bibfnamefont {G.}~\bibnamefont
  {Crowder}}, \bibinfo {author} {\bibfnamefont {L.}~\bibnamefont {Ramunno}},\
  and\ \bibinfo {author} {\bibfnamefont {S.}~\bibnamefont {Hughes}},\
  }\bibfield  {title} {\bibinfo {title} {Quantum trajectory theory and
  simulations of nonlinear spectra and multiphoton effects in waveguide-{QED}
  systems with a time-delayed coherent feedback},\ }\href
  {https://doi.org/10.1103/PhysRevA.106.013714} {\bibfield  {journal} {\bibinfo
   {journal} {Physical Review A}\ }\textbf {\bibinfo {volume} {106}},\ \bibinfo
  {pages} {013714} (\bibinfo {year} {2022})}\BibitemShut {NoStop}%
\bibitem [{\citenamefont {Istrati}\ \emph {et~al.}(2020)\citenamefont
  {Istrati}, \citenamefont {Pilnyak}, \citenamefont {Loredo}, \citenamefont
  {Antón}, \citenamefont {Somaschi}, \citenamefont {Hilaire}, \citenamefont
  {Ollivier}, \citenamefont {Esmann}, \citenamefont {Cohen}, \citenamefont
  {Vidro}, \citenamefont {Millet}, \citenamefont {Lemaître}, \citenamefont
  {Sagnes}, \citenamefont {Harouri}, \citenamefont {Lanco}, \citenamefont
  {Senellart},\ and\ \citenamefont {Eisenberg}}]{Istrati2020}%
  \BibitemOpen
  \bibfield  {author} {\bibinfo {author} {\bibfnamefont {D.}~\bibnamefont
  {Istrati}}, \bibinfo {author} {\bibfnamefont {Y.}~\bibnamefont {Pilnyak}},
  \bibinfo {author} {\bibfnamefont {J.~C.}\ \bibnamefont {Loredo}}, \bibinfo
  {author} {\bibfnamefont {C.}~\bibnamefont {Antón}}, \bibinfo {author}
  {\bibfnamefont {N.}~\bibnamefont {Somaschi}}, \bibinfo {author}
  {\bibfnamefont {P.}~\bibnamefont {Hilaire}}, \bibinfo {author} {\bibfnamefont
  {H.}~\bibnamefont {Ollivier}}, \bibinfo {author} {\bibfnamefont
  {M.}~\bibnamefont {Esmann}}, \bibinfo {author} {\bibfnamefont
  {L.}~\bibnamefont {Cohen}}, \bibinfo {author} {\bibfnamefont
  {L.}~\bibnamefont {Vidro}}, \bibinfo {author} {\bibfnamefont
  {C.}~\bibnamefont {Millet}}, \bibinfo {author} {\bibfnamefont
  {A.}~\bibnamefont {Lemaître}}, \bibinfo {author} {\bibfnamefont
  {I.}~\bibnamefont {Sagnes}}, \bibinfo {author} {\bibfnamefont
  {A.}~\bibnamefont {Harouri}}, \bibinfo {author} {\bibfnamefont
  {L.}~\bibnamefont {Lanco}}, \bibinfo {author} {\bibfnamefont
  {P.}~\bibnamefont {Senellart}},\ and\ \bibinfo {author} {\bibfnamefont
  {H.~S.}\ \bibnamefont {Eisenberg}},\ }\bibfield  {title} {\bibinfo {title}
  {Sequential generation of linear cluster states from a single photon
  emitter},\ }\href {https://doi.org/10.1038/s41467-020-19341-4} {\bibfield
  {journal} {\bibinfo  {journal} {Nature Communications}\ }\textbf {\bibinfo
  {volume} {11}},\ \bibinfo {pages} {5501} (\bibinfo {year}
  {2020})}\BibitemShut {NoStop}%
\bibitem [{\citenamefont {Steindl}\ \emph {et~al.}(2021)\citenamefont
  {Steindl}, \citenamefont {Snijders}, \citenamefont {Westra}, \citenamefont
  {Hissink}, \citenamefont {Iakovlev}, \citenamefont {Polla}, \citenamefont
  {Frey}, \citenamefont {Norman}, \citenamefont {Gossard}, \citenamefont
  {Bowers}, \citenamefont {Bouwmeester},\ and\ \citenamefont
  {Löffler}}]{Steindl2021}%
  \BibitemOpen
  \bibfield  {author} {\bibinfo {author} {\bibfnamefont {P.}~\bibnamefont
  {Steindl}}, \bibinfo {author} {\bibfnamefont {H.}~\bibnamefont {Snijders}},
  \bibinfo {author} {\bibfnamefont {G.}~\bibnamefont {Westra}}, \bibinfo
  {author} {\bibfnamefont {E.}~\bibnamefont {Hissink}}, \bibinfo {author}
  {\bibfnamefont {K.}~\bibnamefont {Iakovlev}}, \bibinfo {author}
  {\bibfnamefont {S.}~\bibnamefont {Polla}}, \bibinfo {author} {\bibfnamefont
  {J.}~\bibnamefont {Frey}}, \bibinfo {author} {\bibfnamefont {J.}~\bibnamefont
  {Norman}}, \bibinfo {author} {\bibfnamefont {A.}~\bibnamefont {Gossard}},
  \bibinfo {author} {\bibfnamefont {J.}~\bibnamefont {Bowers}}, \bibinfo
  {author} {\bibfnamefont {D.}~\bibnamefont {Bouwmeester}},\ and\ \bibinfo
  {author} {\bibfnamefont {W.}~\bibnamefont {Löffler}},\ }\bibfield  {title}
  {\bibinfo {title} {Artificial {Coherent} {States} of {Light} by {Multiphoton}
  {Interference} in a {Single}-{Photon} {Stream}},\ }\href
  {https://doi.org/10.1103/PhysRevLett.126.143601} {\bibfield  {journal}
  {\bibinfo  {journal} {Physical Review Letters}\ }\textbf {\bibinfo {volume}
  {126}},\ \bibinfo {pages} {143601} (\bibinfo {year} {2021})}\BibitemShut
  {NoStop}%
\bibitem [{\citenamefont {Li}\ \emph {et~al.}(2020)\citenamefont {Li},
  \citenamefont {Qin}, \citenamefont {Chen}, \citenamefont {Duan},
  \citenamefont {Yu}, \citenamefont {Huo}, \citenamefont {Höfling},
  \citenamefont {Lu}, \citenamefont {Chen},\ and\ \citenamefont
  {Pan}}]{Li2020}%
  \BibitemOpen
  \bibfield  {author} {\bibinfo {author} {\bibfnamefont {J.-P.}\ \bibnamefont
  {Li}}, \bibinfo {author} {\bibfnamefont {J.}~\bibnamefont {Qin}}, \bibinfo
  {author} {\bibfnamefont {A.}~\bibnamefont {Chen}}, \bibinfo {author}
  {\bibfnamefont {Z.-C.}\ \bibnamefont {Duan}}, \bibinfo {author}
  {\bibfnamefont {Y.}~\bibnamefont {Yu}}, \bibinfo {author} {\bibfnamefont
  {Y.}~\bibnamefont {Huo}}, \bibinfo {author} {\bibfnamefont {S.}~\bibnamefont
  {Höfling}}, \bibinfo {author} {\bibfnamefont {C.-Y.}\ \bibnamefont {Lu}},
  \bibinfo {author} {\bibfnamefont {K.}~\bibnamefont {Chen}},\ and\ \bibinfo
  {author} {\bibfnamefont {J.-W.}\ \bibnamefont {Pan}},\ }\bibfield  {title}
  {\bibinfo {title} {Multiphoton {Graph} {States} from a {Solid}-{State}
  {Single}-{Photon} {Source}},\ }\href
  {https://doi.org/10.1021/acsphotonics.0c00192} {\bibfield  {journal}
  {\bibinfo  {journal} {ACS Photonics}\ }\textbf {\bibinfo {volume} {7}},\
  \bibinfo {pages} {1603} (\bibinfo {year} {2020})}\BibitemShut {NoStop}%
\bibitem [{\citenamefont {Cosacchi}\ \emph {et~al.}(2021)\citenamefont
  {Cosacchi}, \citenamefont {Seidelmann}, \citenamefont {Cygorek},
  \citenamefont {Vagov}, \citenamefont {Reiter},\ and\ \citenamefont
  {Axt}}]{Cosacchi2021}%
  \BibitemOpen
  \bibfield  {author} {\bibinfo {author} {\bibfnamefont {M.}~\bibnamefont
  {Cosacchi}}, \bibinfo {author} {\bibfnamefont {T.}~\bibnamefont
  {Seidelmann}}, \bibinfo {author} {\bibfnamefont {M.}~\bibnamefont {Cygorek}},
  \bibinfo {author} {\bibfnamefont {A.}~\bibnamefont {Vagov}}, \bibinfo
  {author} {\bibfnamefont {D.}~\bibnamefont {Reiter}},\ and\ \bibinfo {author}
  {\bibfnamefont {V.}~\bibnamefont {Axt}},\ }\bibfield  {title} {\bibinfo
  {title} {Accuracy of the {Quantum} {Regression} {Theorem} for {Photon}
  {Emission} from a {Quantum} {Dot}},\ }\href
  {https://doi.org/10.1103/PhysRevLett.127.100402} {\bibfield  {journal}
  {\bibinfo  {journal} {Physical Review Letters}\ }\textbf {\bibinfo {volume}
  {127}},\ \bibinfo {pages} {100402} (\bibinfo {year} {2021})}\BibitemShut
  {NoStop}%
\bibitem [{\citenamefont {Richter}\ and\ \citenamefont
  {Hughes}(2022)}]{Richter2022}%
  \BibitemOpen
  \bibfield  {author} {\bibinfo {author} {\bibfnamefont {M.}~\bibnamefont
  {Richter}}\ and\ \bibinfo {author} {\bibfnamefont {S.}~\bibnamefont
  {Hughes}},\ }\bibfield  {title} {\bibinfo {title} {Enhanced {TEMPO}
  {Algorithm} for {Quantum} {Path} {Integrals} with {Off}-{Diagonal}
  {System}-{Bath} {Coupling}: {Applications} to {Photonic} {Quantum}
  {Networks}},\ }\href {https://doi.org/10.1103/PhysRevLett.128.167403}
  {\bibfield  {journal} {\bibinfo  {journal} {Physical Review Letters}\
  }\textbf {\bibinfo {volume} {128}},\ \bibinfo {pages} {167403} (\bibinfo
  {year} {2022})}\BibitemShut {NoStop}%
\bibitem [{\citenamefont {Gardiner}\ and\ \citenamefont
  {Zoller}(2000)}]{QuantumNoise}%
  \BibitemOpen
  \bibfield  {author} {\bibinfo {author} {\bibfnamefont {C.}~\bibnamefont
  {Gardiner}}\ and\ \bibinfo {author} {\bibfnamefont {P.}~\bibnamefont
  {Zoller}},\ }\href@noop {} {\emph {\bibinfo {title} {Quantum Noise: A
  Handbook of Markovian and Non-Markovian Quantum Stochastic Methods with
  Applications to Quantum Optics}}}\ (\bibinfo  {publisher} {Springer-Verlag},\
  \bibinfo {address} {Berlin},\ \bibinfo {year} {2000})\BibitemShut {NoStop}%
\bibitem [{\citenamefont {Stenius}\ and\ \citenamefont
  {Imamoglu}(1996)}]{Stenius1996}%
  \BibitemOpen
  \bibfield  {author} {\bibinfo {author} {\bibfnamefont {P.}~\bibnamefont
  {Stenius}}\ and\ \bibinfo {author} {\bibfnamefont {A.}~\bibnamefont
  {Imamoglu}},\ }\bibfield  {title} {\bibinfo {title} {Stochastic wavefunction
  methods beyond the {Born} - {Markov} and rotating-wave approximations},\
  }\href {https://doi.org/10.1088/1355-5111/8/1/021} {\bibfield  {journal}
  {\bibinfo  {journal} {Quantum and Semiclassical Optics: Journal of the
  European Optical Society Part B}\ }\textbf {\bibinfo {volume} {8}},\ \bibinfo
  {pages} {283} (\bibinfo {year} {1996})}\BibitemShut {NoStop}%
\bibitem [{\citenamefont {Chruściński}\ and\ \citenamefont
  {Kossakowski}(2010)}]{Chruscinski2010}%
  \BibitemOpen
  \bibfield  {author} {\bibinfo {author} {\bibfnamefont {D.}~\bibnamefont
  {Chruściński}}\ and\ \bibinfo {author} {\bibfnamefont {A.}~\bibnamefont
  {Kossakowski}},\ }\bibfield  {title} {\bibinfo {title} {Non-{Markovian}
  {Quantum} {Dynamics}: {Local} versus {Nonlocal}},\ }\href
  {https://doi.org/10.1103/PhysRevLett.104.070406} {\bibfield  {journal}
  {\bibinfo  {journal} {Physical Review Letters}\ }\textbf {\bibinfo {volume}
  {104}},\ \bibinfo {pages} {070406} (\bibinfo {year} {2010})}\BibitemShut
  {NoStop}%
\bibitem [{\citenamefont {Tian}\ and\ \citenamefont
  {Carmichael}(1992)}]{Tian1992}%
  \BibitemOpen
  \bibfield  {author} {\bibinfo {author} {\bibfnamefont {L.}~\bibnamefont
  {Tian}}\ and\ \bibinfo {author} {\bibfnamefont {H.~J.}\ \bibnamefont
  {Carmichael}},\ }\bibfield  {title} {\bibinfo {title} {Quantum trajectory
  simulations of two-state behavior in an optical cavity containing one atom},\
  }\href {https://doi.org/10.1103/PhysRevA.46.R6801} {\bibfield  {journal}
  {\bibinfo  {journal} {Physical Review A}\ }\textbf {\bibinfo {volume} {46}},\
  \bibinfo {pages} {R6801} (\bibinfo {year} {1992})}\BibitemShut {NoStop}%
\bibitem [{\citenamefont {Dalibard}\ \emph {et~al.}(1992)\citenamefont
  {Dalibard}, \citenamefont {Castin},\ and\ \citenamefont
  {M\o{}lmer}}]{Dalibard1992}%
  \BibitemOpen
  \bibfield  {author} {\bibinfo {author} {\bibfnamefont {J.}~\bibnamefont
  {Dalibard}}, \bibinfo {author} {\bibfnamefont {Y.}~\bibnamefont {Castin}},\
  and\ \bibinfo {author} {\bibfnamefont {K.}~\bibnamefont {M\o{}lmer}},\
  }\bibfield  {title} {\bibinfo {title} {Wave-function approach to dissipative
  processes in quantum optics},\ }\href
  {https://doi.org/10.1103/PhysRevLett.68.580} {\bibfield  {journal} {\bibinfo
  {journal} {Physical Review Letters}\ }\textbf {\bibinfo {volume} {68}},\
  \bibinfo {pages} {580} (\bibinfo {year} {1992})}\BibitemShut {NoStop}%
\bibitem [{\citenamefont {Makhonin}\ \emph {et~al.}(2014)\citenamefont
  {Makhonin}, \citenamefont {Dixon}, \citenamefont {Coles}, \citenamefont
  {Royall}, \citenamefont {Luxmoore}, \citenamefont {Clarke}, \citenamefont
  {Hugues}, \citenamefont {Skolnick},\ and\ \citenamefont
  {Fox}}]{Makhonin2014}%
  \BibitemOpen
  \bibfield  {author} {\bibinfo {author} {\bibfnamefont {M.~N.}\ \bibnamefont
  {Makhonin}}, \bibinfo {author} {\bibfnamefont {J.~E.}\ \bibnamefont {Dixon}},
  \bibinfo {author} {\bibfnamefont {R.~J.}\ \bibnamefont {Coles}}, \bibinfo
  {author} {\bibfnamefont {B.}~\bibnamefont {Royall}}, \bibinfo {author}
  {\bibfnamefont {I.~J.}\ \bibnamefont {Luxmoore}}, \bibinfo {author}
  {\bibfnamefont {E.}~\bibnamefont {Clarke}}, \bibinfo {author} {\bibfnamefont
  {M.}~\bibnamefont {Hugues}}, \bibinfo {author} {\bibfnamefont {M.~S.}\
  \bibnamefont {Skolnick}},\ and\ \bibinfo {author} {\bibfnamefont {A.~M.}\
  \bibnamefont {Fox}},\ }\bibfield  {title} {\bibinfo {title} {Waveguide
  {Coupled} {Resonance} {Fluorescence} from {On}-{Chip} {Quantum} {Emitter}},\
  }\href {https://doi.org/10.1021/nl5032937} {\bibfield  {journal} {\bibinfo
  {journal} {Nano Letters}\ }\textbf {\bibinfo {volume} {14}},\ \bibinfo
  {pages} {6997} (\bibinfo {year} {2014})}\BibitemShut {NoStop}%
\bibitem [{\citenamefont {Uppu}\ \emph {et~al.}(2020)\citenamefont {Uppu},
  \citenamefont {Pedersen}, \citenamefont {Wang}, \citenamefont {Olesen},
  \citenamefont {Papon}, \citenamefont {Zhou}, \citenamefont {Midolo},
  \citenamefont {Scholz}, \citenamefont {Wieck}, \citenamefont {Ludwig},\ and\
  \citenamefont {Lodahl}}]{Uppu2020}%
  \BibitemOpen
  \bibfield  {author} {\bibinfo {author} {\bibfnamefont {R.}~\bibnamefont
  {Uppu}}, \bibinfo {author} {\bibfnamefont {F.~T.}\ \bibnamefont {Pedersen}},
  \bibinfo {author} {\bibfnamefont {Y.}~\bibnamefont {Wang}}, \bibinfo {author}
  {\bibfnamefont {C.~T.}\ \bibnamefont {Olesen}}, \bibinfo {author}
  {\bibfnamefont {C.}~\bibnamefont {Papon}}, \bibinfo {author} {\bibfnamefont
  {X.}~\bibnamefont {Zhou}}, \bibinfo {author} {\bibfnamefont {L.}~\bibnamefont
  {Midolo}}, \bibinfo {author} {\bibfnamefont {S.}~\bibnamefont {Scholz}},
  \bibinfo {author} {\bibfnamefont {A.~D.}\ \bibnamefont {Wieck}}, \bibinfo
  {author} {\bibfnamefont {A.}~\bibnamefont {Ludwig}},\ and\ \bibinfo {author}
  {\bibfnamefont {P.}~\bibnamefont {Lodahl}},\ }\bibfield  {title} {\bibinfo
  {title} {Scalable integrated single-photon source},\ }\href
  {https://doi.org/10.1126/sciadv.abc8268} {\bibfield  {journal} {\bibinfo
  {journal} {Science Advances}\ }\textbf {\bibinfo {volume} {6}},\ \bibinfo
  {pages} {eabc8268} (\bibinfo {year} {2020})}\BibitemShut {NoStop}%
\bibitem [{\citenamefont {Dusanowski}\ \emph {et~al.}(2022)\citenamefont
  {Dusanowski}, \citenamefont {Gustin}, \citenamefont {Hughes}, \citenamefont
  {Schneider},\ and\ \citenamefont {H\"{o}fling}}]{Dusanowski2022}%
  \BibitemOpen
  \bibfield  {author} {\bibinfo {author} {\bibfnamefont {L.}~\bibnamefont
  {Dusanowski}}, \bibinfo {author} {\bibfnamefont {C.}~\bibnamefont {Gustin}},
  \bibinfo {author} {\bibfnamefont {S.}~\bibnamefont {Hughes}}, \bibinfo
  {author} {\bibfnamefont {C.}~\bibnamefont {Schneider}},\ and\ \bibinfo
  {author} {\bibfnamefont {S.}~\bibnamefont {H\"{o}fling}},\ }\bibfield
  {title} {\bibinfo {title} {All-{Optical} {Tuning} of {Indistinguishable}
  {Single} {Photons} {Generated} in {Three}-{Level} {Quantum} {Systems}},\
  }\href {https://doi.org/10.1021/acs.nanolett.1c04700} {\bibfield  {journal}
  {\bibinfo  {journal} {Nano Letters}\ }\textbf {\bibinfo {volume} {22}},\
  \bibinfo {pages} {3562} (\bibinfo {year} {2022})}\BibitemShut {NoStop}%
\bibitem [{sup()}]{supp}%
  \BibitemOpen
  \href@noop {} {}\bibinfo {howpublished}
  {\url{URL_will_be_inserted_by_publisher}}\BibitemShut {NoStop}%
\bibitem [{\citenamefont {Brown}\ and\ \citenamefont
  {Twiss}(1956)}]{Hanbury1956}%
  \BibitemOpen
  \bibfield  {author} {\bibinfo {author} {\bibfnamefont {R.~H.}\ \bibnamefont
  {Brown}}\ and\ \bibinfo {author} {\bibfnamefont {R.~Q.}\ \bibnamefont
  {Twiss}},\ }\bibfield  {title} {\bibinfo {title} {Correlation between
  {Photons} in two {Coherent} {Beams} of {Light}},\ }\href
  {https://doi.org/10.1038/177027a0} {\bibfield  {journal} {\bibinfo  {journal}
  {Nature}\ }\textbf {\bibinfo {volume} {177}},\ \bibinfo {pages} {27}
  (\bibinfo {year} {1956})}\BibitemShut {NoStop}%
\bibitem [{\citenamefont {Hong}\ \emph {et~al.}(1987)\citenamefont {Hong},
  \citenamefont {Ou},\ and\ \citenamefont {Mandel}}]{Hong1987}%
  \BibitemOpen
  \bibfield  {author} {\bibinfo {author} {\bibfnamefont {C.~K.}\ \bibnamefont
  {Hong}}, \bibinfo {author} {\bibfnamefont {Z.~Y.}\ \bibnamefont {Ou}},\ and\
  \bibinfo {author} {\bibfnamefont {L.}~\bibnamefont {Mandel}},\ }\bibfield
  {title} {\bibinfo {title} {Measurement of subpicosecond time intervals
  between two photons by interference},\ }\href
  {https://doi.org/10.1103/PhysRevLett.59.2044} {\bibfield  {journal} {\bibinfo
   {journal} {Physical Review Letters}\ }\textbf {\bibinfo {volume} {59}},\
  \bibinfo {pages} {2044} (\bibinfo {year} {1987})}\BibitemShut {NoStop}%
\bibitem [{\citenamefont {Schnauber}\ \emph {et~al.}(2019)\citenamefont
  {Schnauber}, \citenamefont {Singh}, \citenamefont {Schall}, \citenamefont
  {Park}, \citenamefont {Song}, \citenamefont {Rodt}, \citenamefont
  {Srinivasan}, \citenamefont {Reitzenstein},\ and\ \citenamefont
  {Davanco}}]{Schnauber2019}%
  \BibitemOpen
  \bibfield  {author} {\bibinfo {author} {\bibfnamefont {P.}~\bibnamefont
  {Schnauber}}, \bibinfo {author} {\bibfnamefont {A.}~\bibnamefont {Singh}},
  \bibinfo {author} {\bibfnamefont {J.}~\bibnamefont {Schall}}, \bibinfo
  {author} {\bibfnamefont {S.~I.}\ \bibnamefont {Park}}, \bibinfo {author}
  {\bibfnamefont {J.~D.}\ \bibnamefont {Song}}, \bibinfo {author}
  {\bibfnamefont {S.}~\bibnamefont {Rodt}}, \bibinfo {author} {\bibfnamefont
  {K.}~\bibnamefont {Srinivasan}}, \bibinfo {author} {\bibfnamefont
  {S.}~\bibnamefont {Reitzenstein}},\ and\ \bibinfo {author} {\bibfnamefont
  {M.}~\bibnamefont {Davanco}},\ }\bibfield  {title} {\bibinfo {title}
  {Indistinguishable {Photons} from {Deterministically} {Integrated} {Single}
  {Quantum} {Dots} in {Heterogeneous} {GaAs}/{Si3N4} {Quantum} {Photonic}
  {Circuits}},\ }\href {https://doi.org/10.1021/acs.nanolett.9b02758}
  {\bibfield  {journal} {\bibinfo  {journal} {Nano Letters}\ }\textbf {\bibinfo
  {volume} {19}},\ \bibinfo {pages} {7164} (\bibinfo {year}
  {2019})}\BibitemShut {NoStop}%
\bibitem [{\citenamefont {Gustin}\ \emph {et~al.}(2018)\citenamefont {Gustin},
  \citenamefont {Manson},\ and\ \citenamefont {Hughes}}]{Gustin2018}%
  \BibitemOpen
  \bibfield  {author} {\bibinfo {author} {\bibfnamefont {C.}~\bibnamefont
  {Gustin}}, \bibinfo {author} {\bibfnamefont {R.}~\bibnamefont {Manson}},\
  and\ \bibinfo {author} {\bibfnamefont {S.}~\bibnamefont {Hughes}},\
  }\bibfield  {title} {\bibinfo {title} {Spectral asymmetries in the resonance
  fluorescence of two-level systems under pulsed excitation},\ }\href
  {https://doi.org/10.1364/OL.43.000779} {\bibfield  {journal} {\bibinfo
  {journal} {Optics Letters}\ }\textbf {\bibinfo {volume} {43}},\ \bibinfo
  {pages} {779} (\bibinfo {year} {2018})}\BibitemShut {NoStop}%
\bibitem [{\citenamefont {Mnaymneh}\ \emph {et~al.}(2019)\citenamefont
  {Mnaymneh}, \citenamefont {Dalacu}, \citenamefont {McKee}, \citenamefont
  {Lapointe}, \citenamefont {Haffouz}, \citenamefont {Weber}, \citenamefont
  {Northeast}, \citenamefont {Poole}, \citenamefont {Aers},\ and\ \citenamefont
  {Williams}}]{2019Dan}%
  \BibitemOpen
  \bibfield  {author} {\bibinfo {author} {\bibfnamefont {K.}~\bibnamefont
  {Mnaymneh}}, \bibinfo {author} {\bibfnamefont {D.}~\bibnamefont {Dalacu}},
  \bibinfo {author} {\bibfnamefont {J.}~\bibnamefont {McKee}}, \bibinfo
  {author} {\bibfnamefont {J.}~\bibnamefont {Lapointe}}, \bibinfo {author}
  {\bibfnamefont {S.}~\bibnamefont {Haffouz}}, \bibinfo {author} {\bibfnamefont
  {J.~F.}\ \bibnamefont {Weber}}, \bibinfo {author} {\bibfnamefont {D.~B.}\
  \bibnamefont {Northeast}}, \bibinfo {author} {\bibfnamefont {P.~J.}\
  \bibnamefont {Poole}}, \bibinfo {author} {\bibfnamefont {G.~C.}\ \bibnamefont
  {Aers}},\ and\ \bibinfo {author} {\bibfnamefont {R.~L.}\ \bibnamefont
  {Williams}},\ }\bibfield  {title} {\bibinfo {title} {On-chip integration of
  single photon sources via evanescent coupling of tapered nanowires to {SiN}
  waveguides},\ }\href {https://doi.org/10.1002/qute.201900021} {\bibfield
  {journal} {\bibinfo  {journal} {Advanced Quantum Technologies}\ }\textbf
  {\bibinfo {volume} {3}},\ \bibinfo {pages} {1900021} (\bibinfo {year}
  {2019})}\BibitemShut {NoStop}%
\end{thebibliography}%

\end{document}